The ubiquitous digital file: a review of file management research

Jesse David Dinneen

School of Information Management, Victoria University of Wellington

PO Box 600, Wellington, New Zealand 6140

jesse.dinneen@vuw.ac.nz

+64 4 463 6916

Corresponding author

Charles-Antoine Julien

School of Information Studies, McGill University

3661 Peel St., Montreal, Quebec, Canada, H3A 1X1

charles.julien@mcgill.ca

Abstract

Computer users spend time every day interacting with digital files and folders, including downloading, moving, naming, navigating to, searching for, sharing, and deleting them. Such *file management* has been the focus of many studies across various fields, but has not been explicitly acknowledged nor made the focus of dedicated review. In this paper we present the first dedicated review of this topic and its research, synthesising more than 230 publications from various research domains to establish what is known and what remains to be investigated, particularly by examining the common motivations, methods, and findings evinced by the previously furcate body of work. We find three typical research motivations in the literature reviewed: understanding how and why users store, organise, retrieve, and share files and folders, understanding factors that determine their behaviour, and attempting to improve the user experience through novel interfaces and information services. Relevant conceptual frameworks and approaches to designing and testing systems are described, and open research challenges and the significance for other research areas are discussed. We conclude that file management is a ubiquitous, challenging, and relatively unsupported activity that invites and has received attention from several disciplines and has broad importance for topics across information science.

The ubiquitous digital file: a review of file management research

## Introduction

Computer users spend time every day interacting with digital files and folders, including creating, downloading, naming, moving, saving, copying, reviewing, navigating, searching for, sharing, and deleting them. This activity, called *file management* (FM), is so fundamental and common to knowledge work specifically and modern computer use generally that one could consider a citizen of the information age who never has cause to interact with files to be exceptional. FM is also difficult, personal, deeply psychological in nature (Lansdale, 1988), and increasingly complex, as additions and improvements to FM like desktop search, tagging functions, networked storage, and cloud services have expanded the number possible user interactions and challenges. For example, users can keep files locally, remotely, and in the cloud, synchronise them across devices, organise them as files or as format-specific collections using local and Web-based applications, and navigate and search through them in multiple ways, and they can do all of this as individual users or in collaboration with others. FM is therefore one of the most central activities involved in using a computer, and is thus a fundamental aspect of living in the information society.

FM can be supported by personal information management (PIM) systems and software, especially so if their design is informed by an understanding of the behaviour that users exhibit and its determinant factors. Many studies have worked towards improving this understanding and implementing its lessons in improved systems, and yet this literature and their subject have not been formally acknowledged, reviewed, summarised, synthesised, or reflected upon in a dedicated review. It would aid work in this area to explicate what is known about the phenomenon, how it can be investigated, which aspects could or should be explored next, and how the study of FM can benefit from and contribute to various scholarly disciplines. The goal of this paper is therefore to provide a review of the relevant literature, and in doing so to demarcate the body of scholarly work about file management and understand the current state and limits of its knowledge.

As PIM is the research area most closely related to FM, with FM possibly constituting a subset of PIM research (i.e., with a particular focus on files and folders), many works of PIM are included in this review. FM research is the nonetheless the review target; more precise scope and inclusion criteria are provided in the review methodology, and the relationship of FM to PIM and other areas is explicated in greater depth in the discussion section. A description of the composition of the review follows. In the remainder of the introduction we provide a history and context of files leading to definitions and necessary background information, and then detail our review methodology (goals, literature sources, keywords, etc). The results of our review begin with a detailed synthesis of the motivations of FM research, comprised of understanding user behaviour (storing, organising, retrieving, sharing), understanding internal and external factors determining FM, and designing augmented, novel, or alternative FM systems. We then review the theory, methods, and frameworks seen in the FM literature reviewed, before turning to discuss the review results as a whole. The discussion is composed of two parts, beginning with an explication of the overlap and therefore the importance of FM research to related areas within and beyond the info sciences, wherein we suggest particular topics for future research. In the second half of the discussion we comment on particular challenges awaiting future FM research, including those relevant to each of the areas reviewed (e.g., improved systems, theory and methods, etc), with subsections dedicated to mobile FM and the future of files and FM systems. In the conclusion we summarise the review.

**History and context of files**

The word *file* can have multiple senses related to computer files (Harper et al., 2013), but the sense used in the reviewed literature and adopted here refers to what is perhaps the most common: representations of digital content stored in file systems and presented to users through the metaphor of a paper file (e.g., a physical document, not to be confused with the British use of file to mean *a folder*). In simple cases, files are used to represent, for example, a document, an image, some audio data, an executable program, a database, an ongoing session in some software, or an archive of any of such files. Folders extend the file metaphor to provide categorised access to files and to more folders by containing them, and are presented to the user as though arranged in a spatial hierarchy that starts at a common *root* folder and may contain a minimal default folder structure. Though the term *directory* is often used interchangeably for *folder* (and is the conventional term in some operating systems), in this manuscript we use *folder* to refer to what the user sees or otherwise experiences (e.g., what they see in their file manager, what they move and rename) and *directory* when referring to applications' view of or interaction with locations within the file system.

The focus of this review is thus research into how users interact with digital files as described above, including actions such as the following, though additional relevant actions are conceivable and may become common in the future: creating, naming, renaming, downloading, uploading, attaching, copying, organising, cutting, pasting, tagging, linking via symlink or shortcut, navigating, searching, deleting, and restoring. Such actions are done in contexts that may be occupational (e.g., personally managing company files), personal (e.g., maintaining files for personal use), or both. In other words, what concerns this review is what is seen in file management software, including the common dedicated graphical interfaces (e.g., File Explorer and Finder), command line interfaces, and applications' file open and save dialogues, but not in related contexts where digital items only resemble files; for example, emails presented to a user in a Web-based mail client, though likely sortable into folders, are not the focus of this manuscript as such until they are downloaded and stored as files, for example, by a user saving them to their laptop as eml files. Digital items may also be viewed and managed with items of the same format in particular applications, for example as a collection of songs in iTunes or photos in a photo viewer. Although these items likely also exist as files, we do not regard managing the items within the format-specific application as file management, at least for the purpose of this review, but because such contexts are highly relevant to FM, we address them later in this manuscript.

Today's metaphor for digital content as files organised in folders dates back to the 1960s (Corbató, Merwin-Daggett, & Daley, 1962) and reflects the *memex* envisioned by Vannevar Bush, 1945 by providing a 'private file and library... in which an individual stores... books, records, and communications' (§6), albeit without the associative access Bush had hoped for. Perhaps as a consequence of this utility, the file metaphor has been successful and indeed pervasive in computing for over 40 years (Harper et al., 2013). Although this metaphor for digital content has been questioned (Halasz & Moran, 1982) and warrants critical reflection and refinement (Harper et al., 2013), it is one of the oldest in computing, is widely used (e.g., is used in Windows, Mac OS, GNU/Linux, BSD, Solaris, Android, iOS, Windows Mobile...), and is currently without a serious alternative. However, operating systems (OSes) store, handle, represent, and manage files and folders in a related but different way to how they are presented to, handled, and managed by the typical user (Harter, Dragga, Vaughn, Arpaci-Dusseau, & Arpaci-Dusseau, 2012); for example, in POSIX systems (Mac OS, GNU/Linux), folders (directories) are actually files, devices are represented as files, and the OS and its applications may read and write to a file many times while the user is simply viewing it. This review focuses primarily on files and folders as they are presented to the user, but user-file interaction is of concern as much to those designing file systems as it is to those seeking to understand users' behaviour and improve the relevant software and interfaces.

The original method for performing file management was to enter commands, like *mv* for moving and *cp* for copying, into a prompt (i.e., command line interface). This method persists today, though given the popularity of graphical desktop environments and windowed applications it is likely that most file management is done in graphical file manager applications and dialogues initiated, for example, when opening a file in an application, to directly manipulate file and folder icons. In Microsoft and Apple's current desktop OSes, graphical software for managing files is provided by default (namely, File Explorer and Finder, respectively), and while alternative file managers are available, each with different features and views of files, it is currently unclear to what extent users install or are aware of these.

Regardless of the operating system, users likely spend much time performing the actions described above; so far as we know the exact time spent managing files per day or year has never been calculated for an individual or collectively, but given the most recent estimate from the US Census Bureau is that 78.5% of all households have at least one desktop or laptop computer (File & Ryan, 2014), it is reasonable to assume the aggregate time spent interacting with files is considerable. As we detail below, the phenomenon of FM has received considerable attention from various fields of study, but the resulting body of research has never before been explicitly identified, acknowledged, or synthesised. We next describe our methodology for reviewing the relevant literature, and then proceed to the review.

## Review methodology

The goal of this review is to demarcate the body of scholarly work about the management of digital files and folders in common computing environments (i.e., desktops, laptops, tablets, and mobile phones) and understand the current state of knowledge about the immediately relevant phenomena identified in such work. We pursued this goal by identifying and searching scholarly research databases (e.g., Web of Science, Scopus, Google Scholar...) that index journal articles and conference proceedings dealing with, for example, personal information management, human-computer interaction (e.g., proceedings of ACM conferences), desktop interface design, information behaviour, information science (e.g., *Information Research* and *JASIST*), software development, and personal digital archiving. We searched with keywords including *personal information management*, *file management*, *file system*, *desktop management*, *folder organisation*, and *file retrieval* and various additional permutations. We then scanned the manuscripts' references to identify additional relevant articles and proceedings (i.e., citation pearl growing; Ramer, 2005). We included manuscripts describing information management, personal or otherwise, or general computer use where the management or presentation of files were primary topics of the work, excluding those that did not contain such content. Although some works discussed may be associated with patents, patents were not explicitly searched and included. As the literature to be reviewed could (and indeed did) come from various disciplines, we attempted to include any manuscripts describing cognate and relevant concepts with varying terminology (i.e., performed knowledge translation when possible). We also included in our review any tangentially-related works if they commented on or helped to elucidate trends or topics seen in the included literature.

The result of our literature search was over 200 manuscripts with publications dates from 1960 to 2018, including reports of quantitative and qualitative empirical studies, the development of novel systems at various stages of completion, opinion pieces (e.g., reflection on interface design), and reviews of PIM. We are not aware of any prior, dedicated review of FM (i.e., beyond a subsection in a PIM review or targeted literature review section in an FM study). Table 1 summarises characteristics of the review, its procedure, and its outcomes. Although the works reviewed here are numerous, we accept as a limitation of our review that the fundamental and ubiquitous nature of digital files means that potentially many additional studies exploring cognate issues with differing terminology may not have been seen, despite our best efforts, and thus would be missing from our review. We analysed the included manuscripts to capture common themes, such as motivations and findings, concepts and methods, limitations, and directions identified for future research. The results of our analyses are discussed in turn.

| Review characteristic | Details |
| --- | --- |
| Goal | Identify and review work investigating the management of digital files (i.e., file management) |
| Databases searched | Web of Science, Scopus, Google Scholar, ProQuest, SciTech Premium Collection, Library & Information Science Abstracts (LISA), Applied Social Sciences Index & Abstracts (ASSIA) |
| Keywords used | personal information management, file management, folder management, folder organisation, file system, desktop management, file retrieval, file search, file sharing... |
| Additional strategies | Citation pearl growing, knowledge translation |
| Formats identified | Journal articles, conference proceedings, books and chapters, workshop reports, technical reports |
| Styles identified | User studies, system development and testing papers, reviews, editorials |
| Disciplines identified | personal information management, human-computer interaction, interface design, information behaviour, information science, software development, information retrieval, information visualisation, personal archiving |
| Journals | *ACM Transactions on (1) Information Systems, (2) Computer-Human Interaction (3) Storage; Annual Review of Information Science and Technology; Archivaria; Computer; Human-Computer Interaction; Information Research; Interacting with Computers; International Journal of Man-Machine Studies; Journal of the Association for Information Science and Technology; Journal of Documentation; Library Hi-Tech; Operating Systems Review; Performance Evaluation; Scientific Reports; Software: Practice and Experience...* |
| Proceedings of | Annual Meeting of the Association for Information Science and Technology (ASIST); ACM Conference on (1) Computer-Supported Cooperative Work (CSCW), (2) Human Factors in Computing Systems (CHI), (3) Hypertext and Media (HT), (4) Intelligent User Interfaces (IUI), (5) Information Retrieval (IR), (6) Information Visualisation (IV), (7) Supporting Group Work (GROUP), (8) User Interface Software and Technology (UIST); ASLIB Association for Information Management, iConference, International Conference on Knowledge Management (ICKM), I-SEMANTICS, Joint Conference on Digital Libraries (JCDL)... |
| Exclusion criteria used | Management or presentation of files or similar concepts not a primary topic of the work *except* where work provides commentary on or aids discussion of relevant trends or topics |
| Works included | ~230 |

**Table 1**

*A summary of the review procedure used and its outcomes. Ellipses indicate partial lists.*

## Motivations of file management research

In this section we present FM literature showing three common research motivations: understanding user behaviour (or *what* users do, including when and where they do it), understanding the individual differences and external factors relevant to this behaviour (or *why* they do it), and aiming to improve file management systems and software (or *how* to better support users). Each motivation is discussed and presented with tables summarising the relevant literature, and the theoretical and conceptual frameworks and methodologies employed across the studies are examined in the following section. Two considerations should be kept in mind about our presentation of FM research in this section. First, the works presented span decades during which the nature of computing has evolved, and thus to some extent FM has evolved and users' behaviour likely has as well. It is nevertheless sometimes useful to present together the results of studies from disparate eras of computing; readers should therefore take care when considering the overall results of such studies. Second, as our review revealed that studies of mobile FM are still quite rare (i.e., there is not yet a relevant trend in the literature), we largely omit such work from this section and instead discuss it below in the section on future research directions.

### Understanding user behaviour

Many works have sought to understand the behaviour users exhibit and challenges they face while managing digital files, often following the spirit of research from the 1980s examining how individuals manage paper documents (Case, 1986; I. Cole, 1982; Malone, 1983). This is done under various research topic labels, including but not limited to personal information management, personal digital document management, and personal archiving (or personal digital archiving). Sub-categorisation of this literature reveals four common themes or types of user behaviour examined: file and folder storage (e.g., creating, downloading, naming, managing backups), organisation (e.g., organising the folder structure and categorising files into it), retrieval (e.g., searching, navigating, and tagging), and sharing (e.g., managing shared folders, sending files). The first three of these categories are arguably synonymous with keeping, meta-level, and refinding activities (W. Jones, 2007b), or alternatively with keeping, organising, and exploiting activities (Whittaker, 2011), while sharing activities entail and happen across all other activities. The literature reviewed is reported here along these four behaviour themes, each entailing characterising users' behaviour (or the visible outcomes of their behaviour) and the challenges users face in performing relevant actions. Discussion of users' behaviour within the broader purview of PIM research can be found in the entry on PIM in the *Encyclopaedia of Library and Information Sciences* (W. Jones, Dinneen, Capra, Pérez-Quiñones, & Diekema, 2017).

**Storing.**   Actions done to store files and folders include creating, downloading, naming, moving, copying, backing up, and synchronising (or syncing). Though reports of the number of files users store vary greatly, the number is always large: recent studies have found averages from approximately 4,700 (Hicks, Dong, Palmer, & McAlpine, 2008) to 15,000 files per user (Massey, TenBrook, Tatum, & Whittaker, 2014), with minimums as low as 1,000 (Gonçalves & Jorge, 2003) and maximums as high as 56,994 (Whitham & Cruickshank, 2017). The number of folders stored also varies across studies, for example from 56 (Boardman & Sasse, 2004) to 1,044 (Henderson, 2005), and in a study of one organisation the average increased 370% (from 2,400 to 8,900) over a five year period (Agrawal, Bolosky, Douceur, & Lorch, 2007).

Users' files come from various sources, including the Web (Huvila, Eriksen, Häusner, & Jansson, 2014; W. Jones, Bruce, & Dumais, 2001), external devices (Capra, Vardell, & Brennan, 2014), and peer-to-peer or cloud software (C. Marshall & Tang, 2012), though do not often come from their cell phones, despite the ability to download files to smart phones from the Web (Capra, 2009). Several studies have sought to understand the contents of users' collections, finding for example that document and image files are the most common types kept by students and knowledge workers (Gonçalves & Jorge, 2003; Hicks et al., 2008), and that files may be regarded by users as ephemeral, archived, or current for their intended use (Nardi, Anderson, & Erickson, 1995).

Understanding the challenges of storing so many files is a primary concern of FM studies, and several challenges have been identified. Some challenges are due to the imperfect analogue between the digital desktop and its files with the physical counterparts they are modelled after. For example, some users do not understand the desktop's location in relation to the rest of the accessible disk (Ravasio, Schär, & Krueger, 2004), and for some it is not an attractive place to store files since it is often covered with other windows and does not have the multiple flexible views of its content that the file manager provides (Kaptelinin, 1996). Other challenges are due to the proliferation of digital files – with so many files stored, it is difficult to remember that a file exists when it is needed (W. Jones, Dumais, & Bruce, 2002) – and to the limited support from systems for linking digital documents (i.e., files) to related physical documents (Tayeh & Signer, 2018).

Another aspect to storing files and folders is how they are named, which has drawn attention in FM research as generating meaningful and descriptive names for files and folders helps users find and understand files (Crowder, Marion, & Reilly, 2015), but is also difficult, especially if trying to create names that are concise and unique. In studying file naming behaviour, users have been found to exhibit considerable creativity in file naming (Carroll, 1982), though patterns are identifiable: files are named to display the document they represent, their purpose, a project title, or a relevant creation date or deadline (Hicks et al., 2008), but may also contain characters to facilitate file sorting (W. Jones et al., 2015). Folder names have been found to represent their files' genre, a relevant task, a particular topic meaningful to the user, or a period of time (Chaudhry, Rehman, & Al-Sughair, 2015; Henderson, 2005), and may also represent a priority ranking, their use as storage of information contents, or a combination of these and other themes (Khoo et al., 2007). Folder names may also reflect categories of an external document (e.g., the sections of a curriculum vitae) and, like files, may contain characters to facilitate sorting by name (W. Jones et al., 2015). Beyond alphabetical characters, users also make use of numbers and punctuation such as white space, the underscore, and the hyphen (Gonçalves & Jorge, 2003). The mean length of users' file names may be increasing as the system's limits increase: studies have shown an increase from 6 characters (Carroll, 1982), to 12.6 (Gonçalves & Jorge, 2003), and recently to 18.8 (Fitchett & Cockburn, 2015). Despite all the creativity and possibilities in file naming, duplicated file and folder names are common (Henderson, 2005; Hicks et al., 2008) and are further increased by system-generated folders (Henderson & Srinivasan, 2009); this duplication poses a challenge to retrieving files whether by navigating or searching (e.g., by file name).

The introduction of the cloud and desktop synchronising software, such as Dropbox, has likely changed the nature of users' file storage behaviour, though the nature of this change is still being investigated. Users find synchronisation across devices tedious (Santosa & Wigdor, 2013) and are confused by the cloud and by syncing software: they do not understand what such software is, does, or how it interacts with their local storage or other cloud software (C. C. Marshall, Wobber, Ramasubramanian, & Terry, 2012). They may conceive of it as a file repository, shared repository, personal replication store, shared replication store, and synchronisation mechanism (C. Marshall & Tang, 2012), and try to understand it as it relates to their local storage (Tang, Brubaker, & Marshall, 2013). Users' storage behaviour on the cloud requires further study, as we discuss again below in the context of file sharing behaviour.

A growing portion of studies have examined how and when users back up their files and folders. Though what exactly constitutes a back up is conceived of variedly in the literature, it typically refers to copies of valuable portions of a collection made to provide redundancy or version control, stored on separate physical media of various formats, and not frequently accessed or modified. When backing up their files, people may rely on dedicated back up, sharing, or syncing services such as Dropbox or Apple's Time Capsule to make their back ups (C. Marshall & Tang, 2012). However, some people feel these methods are not reliable or do not fit well into their schedules or operations, and so may initiate and make back ups manually (Capra et al., 2014; Dearman & Pierce, 2008), for example as a consequence of related activities (Kljun, Mariani, & Dix, 2016). Not all users make backups (Kljun et al., 2016), but may nonetheless feel that they should be doing so or doing so more frequently (Kearns, Frey, Tomer, & Alman, 2014).

**Organising.** Organising actions include renaming, creating subfolders, creating shortcuts, symlinks or hard links, filing (e.g., downloading and then moving to a more permanent location), copying directly or by pasting, moving directly or by cut and paste, and deleting (Oh, 2012b). This is typically done using the folder hierarchy, and with various motivations, including giving the user a place for files to persist (Whitham & Cruickshank, 2017) and to help them make sense of, summarise, group, and maintain an overview of the files (W. Jones, Phuwanartnurak, Gill, & Bruce, 2005; Ravasio et al., 2004; Whitham & Cruickshank, 2017), which in turn aids memory about the files organised so that they may be retrieved later (Whitham & Cruickshank, 2017; Xie, Sonnenwald, & Fulton, 2015). The file manager software installed on virtually every computer aims to facilitate this process, but with uncertain results: some users have reported that the locations of OS-provided default folders are confusing and that the system-ascribed metadata was not useful for understanding their collections (Ravasio et al., 2004), while others make distinct use of the desktop, default folders, and secondary folders (Paré, 2011). Despite this confusion, some users do indeed store files in default folders, including files that are *active* or currently being frequently accessed (Bergman, Whittaker, Sanderson, Nachmias, & Ramamoorthy, 2010), in locations such as *My Documents* and on the desktop (Khoo et al., 2007); one study found use of the default folders among users at one organisation grew over five years, accounting for 40% of all files (Agrawal et al., 2007).

Users also create, arrange, and remove folders (e.g., when revising their folder structure to accommodate files that were previously difficult to categorise; Oh & Belkin, 2014) in locations beyond the default folders, such as the root of their hard drives (Ravasio et al., 2004) or at the root of their home folders. A developing line of research into the process of organising files has begun to identify discrete stages in FM organisation, including initiation; identification; temporary categorisation; examination and comparison; selection, modification, and creation; and categorisation (Oh, 2012a, 2013), and has begun to extend characterisations of users' organisational styles (e.g., filers and pilers as first suggested by Malone, 1983) with descriptors like fuzzy, rigid, or flexible (Oh, 2017).

In performing organisation actions, users determine the shape of the overall folder tree with which they interact, and that shape can be described by measuring properties of the tree like its height, depth (the maximum number of steps taken when navigating into consecutive subfolders), and consistency (deviation in shape among its main branches).

Such measurements comprise a tree topography, or quantitative description of how people organise their digital items, from which descriptions of particular aspects of user behaviour can be identified, measured, and described (e.g., two thousand files outside of default locations but within the user's folders imply they have chosen to create or acquire and store as many). Descriptions such as these have been provided in many studies, including in those using quantitative data to complement and give scale to qualitative findings. Among disparate contexts, participant groups, and file system measures used in previous studies, users' organising behaviour has been found to vary wildly.

As mentioned above, users may be spreading their files across as few as 56 folders or as many as 9000. Figures reflecting the total number of folders are of limited use in analysing the organisation behaviour, however, as folders can contain any number of files; they can, for example, contain many files, acting as traditional storage locations, or only other folders, thus acting as a navigation fork (Bergman et al., 2010).

User-created hierarchies may vary greatly in maximum depth; a range from one level of depth (i.e., no subfolders) to sixteen levels deep was found in a single study (Henderson & Srinivasan, 2011). Deeper structures are in part a result of larger collections (Henderson & Srinivasan, 2009), and depth in turn contributes to an increase in file name redundancy (Henderson, 2011) and time required to retrieve files (Bergman et al., 2010). Users may create hierarchies that display consistency among their internal *branches* (Gonçalves & Jorge, 2003) or not (Henderson, 2005). Hierarchies may present the user with many navigation decisions by having a high average *branching factor*, or number of subfolders per folder, of 41.8 (Hicks et al., 2008), or a very low branching factor, for example of only 1.84 (Gonçalves & Jorge, 2003).

The location of files within the folder hierarchy is another object of inquiry in FM studies, as where users put their files later affects how long it takes to retrieve them (Bergman, Gradovitch, Bar-Ilan, & Beyth-Marom, 2013). Users may file every single document despite each classification action being cognitively demanding (Ravasio et al., 2004), or may leave up to 6.5% unfiled (Henderson & Srinivasan, 2011), sometimes called *dumping*, perhaps because they are unsure where to put a file, do not have time to file it, or want it to be easily accessible (e.g., on the desktop; Kamaruddin & Dix, 2010). Depending on the user, filing the average file may mean storing it just two levels down from the root of the tree (Bergman, Whittaker, & Falk, 2014), while others store most files in deeper levels (Hicks et al., 2008). As the number of files in a folder increases, so does the work required to review them all, and although users report creating new subfolders when a folder contains 3-7 items (Ravasio et al., 2004), the average number of files found in folders has ranged from low figures like 0 (Henderson & Srinivasan, 2009) or 4 (H. Zhang &

Hu, 2014) to 12 (Bergman et al., 2010; Henderson & Srinivasan, 2009) or 16 (Gonçalves & Jorge, 2003; Hardof-Jaffe, Hershkovitz, Abu-Kishk, Bergman, & Nachmias, 2009b). Factors that likely influence the number of items in a folder include, among others, the amount of time that has passed since the collection or folder was created (i.e., more time provides more chances to add files; Boardman & Sasse, 2004) and the format of the files (e.g., users may store all their photos in one folder but are less likely to store all documents in one folder; W. Jones et al., 2015). Default folders have been found to have a mean of 19.42 files per folder (Bergman et al., 2010), and so may be fuller than folders in completely user-created branches, perhaps because they are more likely to be used to store frequently accessed documents (e.g., active project files); a complete comparison will require examining and comparing folder structures beyond those housing recently accessed files, including on devices where backups or personal archives are stored.

Studies have conflicting reports of users creating folders without putting files into them: while one study found users typically do not create empty folders (Khoo et al., 2007), another found that most users do, with 8% (mean) of folders being empty (Henderson & Srinivasan, 2009), and higher percentages being reported in other studies, including 14% (Sienknecht, Friedrich, Martinka, & Friedenbach, 1994) to 18% (Douceur & Bolosky, 1999). Empty folders may be made, for example, by putting nothing in them at the point of their creation, perhaps in anticipation of forthcoming projects Kamaruddin, Dix, and Martin, 2006; Khoo et al., 2007, or by not deleting them when the last file or folder is removed. Users thus appear to differ widely regarding most FM actions; what they seem to have in common, however, is a lack of reliance on soft file linking features such as aliases in Mac, shortcuts in Windows, and symlinks in Linux (Gonçalves & Jorge, 2003; Ravasio et al., 2004). Nonetheless, different approaches or strategies to organising have been identified among the varied findings, albeit rather broadly, so that we can describe organisers as: neat or messy (Boardman & Sasse, 2004), prone to saving or deleting (Berlin, Jeffries, O'Day, Paepcke, & Wharton, 1993), and prone to filing or piling (Malone, 1983), extensive filing or single folder filing (Henderson & Srinivasan, 2011), or mixing approaches (Trullemans & Signer, 2014a). To draw conclusions beyond these, studies are needed with commensurable contexts, participant characteristics, file system measures, and results reporting (Dinneen, Odoni, Frissen, & Julien, 2016).

**Retrieving.** Retrieving files and folders may be done to find them for the first time (e.g., in a shared drive) or to return to them (Dumais et al., 2016); returning to an item is also called refinding, and is distinct from simply finding an item again because the user has additional information about their existence and location and thus may have additional retrieval methods available (Capra, Pinney, & Perez-Quinones, 2005). Specifically, retrieving can be done manually, for example by navigating through the folder hierarchy to a file's location, or by searching, for example by file property, keyword, or tag label. Both approaches to retrieval require remembering *something* about the object to be retrieved, such as its location, name, or other properties.

Much FM research has been motivated by understanding navigation and comparing it with search, typically by examining users' behaviour and preferences and their influences. A preference for navigating to files is much more common than a preference for searching (Fitchett & Cockburn, 2015; Song & Ling, 2011), even among users who prefer to search rather than navigate folders when retrieving their emails (W. Jones, Wenning, & Bruce, 2014). There are numerous potential causes for this; users report that they feel desktop search tools are too complicated (Ravasio et al., 2004), the search results are too numerous and not meaningfully ranked (Fitchett & Cockburn, 2015), and that navigating through folders provides important reminding cues about their collections (Barreau & Nardi, 1995).

These reports are reflected in users' behaviour: users perform navigation far more than searching (Bergman, Beyth-Marom, Nachmias, Gradovitch, & Whittaker, 2008; Fitchett & Cockburn, 2015), even when they knew the name of what they were looking for (Teevan, Alvarado, Ackerman, & Karger, 2004), are given improved search engines (Bergman, Beyth-Marom, Nachmias, et al., 2008), or have not made the effort to maintain a highly-structured information organisation (Teevan et al., 2004). Users typically search their files only as a last resort in rare cases when navigation fails (Bergman, Beyth-Marom, Nachmias, et al., 2008; Fitchett & Cockburn, 2015; Nardi et al., 1995), for example when a folder structure has become unfamiliar over time (Čopič Pucihar, Kljun, Mariani, & Dix, 2016; Narayan & Olsson, 2013). The tendency not to search unless it is necessary is likely due, in part, to navigation being easier to perform: it allows users to explicate less of their information need and the folders presented at each step provide additional context to guide the navigation (Teevan et al., 2004). A neuro-cognitive explanation of the relative ease of navigation has been explored in recent studies (summarised by Bergman & Benn, 2018): navigating tasks require less cognitive effort than searching tasks (Bergman, Tene-Rubinstein, & Shalom, 2013), and large portions of the brain dedicated to spatial cognition and used in real world navigation are activated during FM navigation, whereas smaller areas dedicated to linguistic processing are activated during search tasks (Benn et al., 2015).

A third option for retrieving files is to search by tags; tagging provides an alternative to classifying files into folders by allowing users to assign numerous labels to files that can later be searched or browsed and taxing classification and navigation tasks can be avoided, for example deciding which single folder a file should be placed within. Because of its promise and use in Web-based contexts and email, tagging has been studied in those contexts, where navigation was still preferred by users to searching tags (Civan, Jones, Klasnja, & Bruce, 2008), and tagging entailed cognitive load for the user (e.g., when deriving label names; Gao, 2011). This has been reflected in FM research into tagging, where users report being less frustrated with folders than tags, and in the end use folders more than tags (Bergman, Gradovitch, Bar-Ilan, & Beyth-Marom, 2013) even when their reported preference was for tagging and they were provided both systems (Bergman, Gradovitch, Bar-Ilan, & Beyth-Marom, 2013). Experienced users may tag faster than they file (Voit, Andrews, & Slany, 2012b), but rarely apply more than one tag (Bergman, Gradovitch, Bar-Ilan, & Beyth-Marom, 2013), thus losing some of the value of the potential for multiple classification of files. The takeaway is somewhat unclear, as noted by (Bergman, Gradovitch, Bar-Ilan, & Beyth-Marom, 2013): from the findings of many studies of tagging, one can see that both folders and tags are better, worse, and no different than their alternative at any given aspect of retrieval. Therefore, work remains to provide the kind of explanation for tagging-vs-filing behaviour and preferences that has recently been done for searching-vs-filing.

Despite relatively clear indications that users prefer and perform navigation in FM contexts, desktop search has a discrete purpose and tagging shows promise, and therefore search and tagging systems will likely continue to develop in the coming years. Improved FM tools may, for example, usefully integrate search and navigation functions (Fitchett, Cockburn, & Gutwin, 2014; Julien, Asadi, Dinneen, & Shu, 2016), or improve searching capabilities by utilising the extensive metadata that users are more likely to remember (Gonçalves & Jorge, 2008a), such as file provenance (Jensen et al., 2010), file type (Blanc-Brude & Scapin, 2007), and time (Dumais et al., 2016).

**Sharing.**   Interacting with shared files and folders involves sharing them or having them be shared with you, and then performing the usual storage, organisation and retrieval tasks in a way influenced by the fact that they are shared (i.e., co-managing, also known as Group Information Management or GIM). Sharing files may be a relatively simple and singular act, for example when users share files on USB sticks or in email attachments for personal purposes (Capra et al., 2014) or to circumvent institutional access control policies or difficult software interactions (M. L. Johnson, Bellovin, Reeder, & Schechter, 2009). It may, however, be a complex negotiation of a shared information spaces (e.g., a Dropbox folder or company intranet) and organisational needs, which leaves the files fragmented across multiple locations or services (Čopič Pucihar et al., 2016; Tang et al., 2013; A. Voida, Olson, & Olson, 2013).

Various problems arise in shared file management contexts. Some problems are relatively simple and likely easy to fix; for example, some interfaces for shared files (e.g., Google Drive) do not always implement the typical features of file managers, like a dialogue for saving files to a particular location (i.e., instead saving files to an implicit root folder rather than a user-created folder). The problem that arises from this particular example is that users retrieve their files from such spaces less successfully and less quickly, but the issue is successfully resolved by implementing the missing dialogue (Bergman, Whittaker, & Frishman, 2018).

Other issues with shared file management are more complicated. For example, individuals' personal information access strategies break down when managing group information, and people struggle to find files in unfamiliar or unintelligible information structures created by others for their own use (Berlin et al., 1993; Čopič Pucihar et al., 2016). Such difficulties may be due to a lack of *mutual intelligibility*; customisation done to make information structures more meaningful for one person often makes them less accessible or intelligible to others (Dourish, Lamping, & Rodden, 1999). If the management of shared folders is treated with an inclusive approach where nothing is deleted, files become copies across multiple versions and locations (i.e., forked), folders may get messy, users may run out of hard drive space (Capra et al., 2014), and users may forget what is shared and thus forget to maintain or cease sharing it (Khan, Hyun, Kanich, & Ur, 2018). If, however, users intend to tidy the shared space, it may be unclear to them who owns any given file (H. Zhang & Twidale, 2012), and they will typically face a lack of policy and direction regarding deletion, naming, and organisation (Capra et al., 2014).

Making changes to item names and moving or deleting items entails changes that other users will may be unaware of and possibly frustrated with (Čopič Pucihar et al., 2016; H. Zhang & Twidale, 2012). In turn, retrieval in shared folders is more time-consuming and prone to error than retrieval from one's own folders, and users may prefer peer-to-peer sharing acts to co-managing a shared information space (Bergman et al., 2014) so long as retrieving co-managed files appears more complicated than retrieving one's own. These problems sometimes lead to the establishment of explicit conventions, strategies, and even tools (Massey, Lennig, & Whittaker, 2014) for managing the shared space, which may tie usefully into other aspects of group work (e.g., establishing a division of labour with the files' contents; Wulf, 1997), but these can be difficult to identify, establish, and follow (Mark & Prinz, 1997). Therefore, implicit and assumed rules often guide users' behaviour (H. Zhang & Twidale, 2012), and thus warrant further study.

Table 2 presents studies that have examined user behaviour, categorised by FM behaviour theme.

| FM theme | Example actions | Studies |
|---|---|---|
| Storing | creating, downloading, filing, naming, backing up files | Barreau (1995), Capra (2009), Capra, Vardell, and Brennan (2014), Carroll (1982), Crowder, Marion, and Reilly (2015), Dearman and Pierce (2008), Gonçalves and Jorge (2003), Henderson (2005), Henderson and Srinivasan (2009), Hicks, Dong, Palmer, and McAlpine (2008), Huvila, Eriksen, Häusner, and Jansson (2014), W. Jones, Bruce, and Dumais (2001), W. Jones et al. (2015), W. Jones, Dumais, and Bruce (2002), Kaptelinin (1996), Kearns, Frey, Tomer, and Alman (2014), Khoo et al. (2007), Kljun, Mariani, and Dix (2016), C. C. Marshall, Wobber, Ramasubramanian, and Terry (2012), C. Marshall and Tang (2012), Nardi, Anderson, and Erickson (1995), Ravasio, Schär, and Krueger (2004), Santosa and Wigdor (2013), Tang, Brubaker, and Marshall (2013), Tayeh and Signer (2018) |
| Organising | creating subfolders, moving and deleting files and folders | Bergman, Whittaker, Sanderson, Nachmias, and Ramamoorthy (2010), Berlin, Jeffries, O'Day, Paepcke, and Wharton (1993), Boardman and Sasse (2004), Chaudhry, Rehman, and Al-Sughair (2015), Gonçalves and Jorge (2003), Hardof-Jaffe, Hershkovitz, Abu-Kishk, Bergman, and Nachmias (2009b), Henderson (2005, 2011), Henderson and Srinivasan (2009, 2011), Hicks, Dong, Palmer, and McAlpine (2008), W. Jones, Phuwanartnurak, Gill, and Bruce (2005), Kamaruddin and Dix (2010), Kamaruddin, Dix, and Martin (2006), Kaptelinin (1996), Malone (1983), Oh (2012a, 2012b, 2013, 2017), Oh and Belkin (2014), Paré (2011), Ravasio, Schär, and Krueger (2004), Trullemans and Signer (2014a), Whitham and Cruickshank (2017), H. Zhang and Hu (2014) |
| Retrieving | navigating, searching, tagging files and folders | Barreau and Nardi (1995), Benn et al. (2015), Bergman, Beyth-Marom, Nachmias, Gradovitch, and Whittaker (2008), Bergman, Gradovitch, Bar-Ilan, and Beyth-Marom (2013), Bergman, Gradovitch, Bar-Ilan, and Beyth-Marom (2013), Bergman, Tene-Rubinstein, and Shalom (2013), Bergman, Whittaker, and Falk (2014), Cutrell (2006), Cutrell, Dumais, and Teevan (2006), Fitchett and Cockburn (2015), Jensen et al. (2010), W. Jones, Wenning, and Bruce (2014), Narayan and Olsson (2013), Nardi, Anderson, and Erickson (1995), Ravasio, Schär, and Krueger (2004), Song and Ling (2011), Teevan, Alvarado, Ackerman, and Karger (2004), Voit, Andrews, and Slany (2011, 2012b) |
| Sharing | sending files, negotiating storage, organisation, retrieval in shared space | Bergman, Whittaker, and Falk (2014), Bergman, Whittaker, and Frishman (2018), Berlin, Jeffries, O'Day, Paepcke, and Wharton (1993), Capra, Vardell, and Brennan (2014), Čopič Pucihar, Kljun, Mariani, and Dix (2016), Dourish, Lamping, and Rodden (1999), M. L. Johnson, Bellovin, Reeder, and Schechter (2009), Khan, Hyun, Kanich, and Ur (2018), Mark and Prinz (1997), Massey, Lennig, and Whittaker (2014), Tang, Brubaker, and Marshall (2013), A. Voida, Olson, and Olson (2013), Wulf (1997), H. Zhang and Twidale (2012) |

**Table 2**

*FM studies seeking to understand user behaviour, presented along common themes.*

## Understanding individual differences and external factors

Understanding user behaviour and supporting it with improved software both entail understanding how users' individual differences and broader contexts could determine their behaviour. The few studies of these factors' roles in FM are discussed below; a review of their roles in standard PIM contexts like email, the Web, and paper documents is provided by Gwizdka and Chignell (2007).

**Individual differences.** Though it is acknowledged that PIM is deeply personal and psychological (Lansdale, 1988), the current knowledge about how individual differences affect users' behaviour is still minimal, especially with regards to FM.

Some of the concern for individual differences in FM contexts has been on spatial cognition, which is appropriate to the file and folder metpahor: folders are represented as being contained within one another and displayed in a two-dimensional space (i.e., represented spatially), users *navigate* through the folder hierarchy space, and users can (and do) arrange files and folders on the desktop as part of their organisational strategy (Ravasio et al., 2004). In virtue of the spatial presentation of files, FM stands to benefit from a large body of work on spatial layout and navigation in computer graphics (c.f.: Chalmers, 1993; Dourish & Chalmers, 1994); due to space constraints we restrict this discussion to individual spatial cognition in FM. An early study of FM found that participants with low spatial ability took twice as long to complete navigation tasks in terminal-based (i.e. text-only, without icons) hierarchical file systems (Vicente, Hayes, & Williges, 1987), although this difference could be partially alleviated with the addition of a simple map (Vicente & Williges, 1988). The terminal-based paradigm for file interaction is no longer the predominant one, and as of yet no work has specifically looked for similar effects in the modern graphical paradigm. It has, however, been noted that users develop preferences towards using either the spatial layout of their folders or patterns in file names when retrieving their files (Krishnan & Jones, 2005), suggesting an active role of spatial ability in modern FM. As discussed above, some work has recently found neuro-cognitive indicators of why spatial cognition plays such a role (Benn et al., 2015), and so future studies may carry out finer-grained investigations of how different FM actions are affected by this role.

Works complementing those discussed above focus on the antecedent to retrieval and navigation, namely, organisation, and the individual differences that may influence it. So far such works have, for example, begun to consider that individual's flexibility in thinking may influence their organisational strategy (Oh, 2017). Notable works have also examined the role of personality style and emotion in FM. Extending into the digital domain work on the influence of personality on the organisation of physical spaces (Gosling, Ko, Mannarelli, & Morris, 2002), one study examined features of file organisation that participants assumed would predict personality (Massey, TenBrook, et al., 2014), and found that conscientious participants were more active organisers, keeping fewer files overall and more files per folder (i.e., fewer folders), but also more files on the desktop. Surprisingly, neuroticism and openness were not correlated with organisational or storage behaviour; future studies examining additional measures of FM may complement these findings. Across two studies examining mood (Massey, 2017) it was found that momentary changes in mood can affect user's FM behaviour, as "sad participants made significantly more folders than happy participants" (p. vii), but that there was no clear relationship between organisation and longer-term trait emotional tendency (e.g., generally happy people didn't have significantly fewer folders or shallower hierarchies than generally unhappy people).

In addition to revealing determining factors that may generalise to other contexts where users organise and retrieve information, studies of individual differences in FM may also discover effective ways to support individuals in FM and similar tasks, for example through detailed user modelling and software personalisation. Further such directions for this line of research are discussed below.

**External factors.**    It is established that external or contextual factors such as occupation, information task, or time are important to understanding the use of paper documents (Kwaznik, 1991) and digital PIM systems (Capra & Perez-Quinones, 2006). This is a concern in FM research as well, but these factors are not yet well understood. For example, the specific effects of occupation are unclear: though participants' occupations have been suggested to be a factor in determining folder naming strategies (Khoo et al., 2007), folder tree height (H. Zhang & Hu, 2014), and folder organisation (Paré, 2011), occupation seems to have no effect on branching factor (Gonçalves & Jorge, 2003), and findings disagree about the effect of occupation on the total number of files stored (Agrawal et al., 2007; Gonçalves & Jorge, 2003; Henderson & Srinivasan, 2009).

The specific effects of occupation may become clearer as they are explored more narrowly. This may include specific occupational traits like regular activities, demands, and patterns and constraints on time spent organising and retrieving information. Notably, the personal or collaborative management of work files is likely determined in part by institutional policy, for example to delete anything older than two years, or keep everything for at least five years; such policy may be followed, thus determining the contents of a file collection as they do with email (M. L. Johnson et al., 2009), or circumvented if employees find them too onerous (M. L. Johnson et al., 2009).

Another external factor of concern in FM research is the tools used to perform FM: the PIM tool adopted for some task enables, restricts, and affects behaviour of the user (Boardman & Sasse, 2004; Fertig, Freeman, & Gelernter, 1996a), as do tools' surrounding software environments (Kaptelinin, 1996) and the hardware they are housed in. In the context of FM, this includes the computer or hardware device, hard disks, file manager software (sometimes called a file browser – the most popular of which are File Explorer in Windows and Finder in Mac OS), windowing environment (if any), and the operating system (OS). For example, limited disk space and even limited cloud storage space can force users to decide what to store locally and what to archive externally (Barreau & Nardi, 1995; Kljun et al., 2016), and the available views onto files provided by the file manager may influence how files are organised and retrieved. Though the exact differences between the software relevant to FM have yet to be thoroughly catalogued – for example, the differing OSes and their respective file manager applications allow, encourage, discourage, and forbid different interactions with files – it has been suggested by ancillary analyses in several studies (Agrawal et al., 2007; Barreau & Nardi, 1995; Massey, TenBrook, et al., 2014) that such differences may affect users' file storage, management, retrieval, and sharing behaviour. For example, an additional finding of Massey, TenBrook, et al. (2014) was that among participants using Windows, conscientiousness was positively correlated with the number of files kept on the desktop, but no such correlation existed for Mac users.

Only one study has explicitly investigated such potential affects (Bergman, Whittaker, Sanderson, Nachmias, & Ramamoorthy, 2012): while participants retrieved files from their own computers, the researchers noted participants' operating system, file manager presentation mode, retrieval times and success rates, and file and folder organisation. Though collecting data only about recently accessed files, they found that Mac users retrieve files faster than Windows users as a result of a differing organisational strategy: they keep more folders close to the root, with fewer files but more subfolders per folder. They also found that the file manager presentation style with which users performed retrievals best was the icons view, regardless of the OS, and therefore suggested that the Windows default should therefore be changed from the details-based view; users rarely change such defaults (Barreau, 1995). This constitutes a good starting point for understanding the effect of the tool on FM behaviour, and future studies may therefore seek to understand the effects of the OS, file manager, and cloud storage software on storage behaviour and additional variables in organisational behaviour exhibited across participants' recent and archived files.

Hardware, too, may affect users' FM behaviour; limited available hard drive space may cause users to save fewer large files or transfer files to the cloud or external physical drives, and users may be less likely to perform intensive FM actions (like navigating deep trees or making backups) when using a laptop (i.e., using a touch pad, relatively small monitor, and small hard drive) than they would be with typical desktop hardware. Few FM studies have touched upon such topics, but the growth of hard drive capacity, and thus of file storage capacity, can be seen over time in the FM literature. For example, in the mid 1990s users had, roughly, only 80 MB to 1.5 GB of storage space (Nardi et al., 1995), but in a study of one work place taking place a decade later, the mean capacity per participant increased from 8 to 46 GB over a five year period (Agrawal et al., 2007). In that study mean hard drive consumption grew from only 3 to 18 GB across five years, suggesting that at least the employees at that organisation are not restricted by hard drive space, but the adoption of faster, smaller solid state hard drives may introduce another factor into this trend.

A table summarising the individual differences and external factors that have been examined for their role in determining FM behaviour are presented in Table 3.

| Group | Factors | Studies |
|---|---|---|
| Individual differences | mood, personality style, spatial cognition and ability, perceived importance of documents | Benn et al. (2015), Bergman and Benn (2018), Kwaznik (1991), Lansdale (1988), Massey (2017), Massey, TenBrook, Tatum, and Whittaker (2014), Oh (2017), Paré (2007), Vicente, Hayes, and Williges (1987), Vicente and Williges (1988) |
| External factors | tool (hardware, OS, FM software), context, information type, time, occupation, task | Agrawal, Bolosky, Douceur, and Lorch (2007), Barreau (1995), Bergman, Whittaker, Sanderson, Nachmias, and Ramamoorthy (2010, 2012), Douceur and Bolosky (1999), Fertig, Freeman, and Gelernter (1996a), Gonçalves and Jorge (2003), Henderson and Srinivasan (2009), W. Jones, Dumais, and Bruce (2002), Kaptelinin (1996), Khoo et al. (2007), Nardi, Anderson, and Erickson (1995), Paré (2011), H. Zhang and Hu (2014) |

**Table 3**
*FM studies seeking to understand individual differences and external factors determining FM behaviour.*

## Improving FM systems

One of the main goals of FM research, as with broader PIM research, is to save users time and effort, and to understand and support their behaviour through improved file management software. There have been many attempts at this, generally either in the form of augmentations to existing FM software or new and alternative metaphors for handling digital content intended to replace some or all of the hierarchical arrangement of files and folders. In both cases the systems are generally purposefully designed, prototypes are built, and these are then tested with live users in semi-natural use or structured experiments (such methodologies are reviewed later in this paper). Although these systems typically have short lives and do not transfer into mainstream use, the novel concepts they develop and evaluate often do eventually trickle into commonly used software (Kljun, Mariani, & Dix, 2015). We review here both augmentations to FM existing software and alternative approaches to managing personal digital content.

**FM software augmentations.** One approach to facilitating file management is to design augmentations to existing FM software to test intuitions about improvements in FM interaction and treat challenges identified in previous studies. By aiming to incrementally improve the current file management paradigm this approach benefits from not overloading users with the task of learning a new system (Bondarenko & Janssen, 2005) or surprising them with unfamiliar metaphors or interfaces (Seebach, 2001).

One motivation in augmenting the file manager is to aid the user when navigating through the folder hierarchy. The oldest of these augmentations improved navigating the folder hierarchy in the command line by providing a map of the hierarchy with the user's current location (Vicente & Williges, 1988), and this was found to enable users with low spatial ability to perform retrieval tasks with the same efficacy as users with high spatial ability. More recent attempts improve graphical navigation, for example by highlighting a path to folders that contain file search matches (Fitchett, Cockburn, & Gutwin, 2013, 2014). Navigation has also been improved by allowing users to de-emphasise files (Bergman, Elyada, Dvir, Vaitzman, & Ami, 2015; Bergman, Tucker, Beyth-Marom, Cutrell, & Whittaker, 2009) and by hiding unused folders (Lee & Bederson, 2003) so that fewer navigation decisions are required during re-finding tasks.

As discussed above, there are instances where re-finding by navigation fails and desktop search may be used as a last resort; several studies have sought to augment the relevant software used in such cases. Most of these have focused on improving general search algorithms and interfaces (B. Cole, 2005; Ghorashi & Jensen, 2012; Kim & Croft, 2010; W. Liu, Rioul, McGrenere, Mackay, & Beaudouin-Lafon, 2018; Sauermann, Bernardi, & Dengel, 2005; Seenu, Rao, & Padma, 2014) or applying semantic search to the desktop (Adrian, Klinkigt, Maus, & Dengel, 2009; Handschuh, Möller, & Groza, 2007; Sauermann et al., 2006), for example by using semantic attributes to enhance search ranking (Chirita, Costache, Nejdl, & Paiu, 2006). Others, however, have sought to support specific search contexts, such as finding similar or duplicate files (Manber, 1994) or supporting search with a more interactive interface and drawing on a detailed file metadata index (G. Liu, Jiang, & Feng, 2017). Further tools for searching across PIM objects beyond files and folders are reviewed by Cutrell (2006).

Several FM software augmentations have been motivated by improving the social and networked aspect of file management by supporting the management of shared and cloud-based files and folders. Some augmentations simplify the users' interactions, for example by providing a unified view of local and cloud folders (W. Jones, Thorsteinson, Thepvongsa, & Garrett, 2016), using content and task analysis to suggest locations for new documents to be placed (Prinz & Zaman, 2005), or unifying synchronisation across a users' devices and across multiple users (C. C. Marshall et al., 2012). Other augmentations have aimed to make the complexity of social file management more intelligible to users, for example by allowing them to review the permissions of all shared files (S. Voida, Edwards, Newman, Grinter, & Ducheneaut, 2006), storing the history of shared files (Whalen, Toms, & Blustein, 2008), and visualising the history and permissions metadata (Rode et al., 2006); these augmentations therefore help to clarify the consequences of users' actions on other users' interactions and on the security of their own digital possessions.

In addition to aiding users in understanding shared files, increased file metadata has been used to try to improve the usefulness of files and their retrieval. So far, this has been done manually, by allowing users to link their files to Web resources (Tayeh, Ebrahimi, & Signer, 2018) and assign annotations and images to their folders (W. Jones, Hou,

Sethanandha, Bi, & Gemmell, 2010; W. Jones, Thorsteinson, et al., 2016), and automatically, by enriching files and folders with content taken from a relevant Web source (He, Li, & Shen, 2013; S. Voida & Greenberg, 2009), or reading file contents to assign them representative icons (Roy, Singh, Chawla, Saxena, & Sinha, 2017). Finally, small but ubiquitous FM actions have not been overlooked, as augmentations have aimed to: make filing new files easier by suggesting locations (Prinz & Zaman, 2005; Sinha & Basu, 2012b), improve file copying tasks by adding a many-to-one feature (Sinha & Basu, 2012c), make moving sensitive files to external destinations more secure alerting users if the files have remained in the original location (i.e., if files have been copied rather than moved; Ishizawa, Andoh, & Nishida, 2010), facilitate planned backups (Cox, Murray, & Noble, 2002), increase the likelihood of retrieving cloud files by prompting users to store them in folders (Bergman et al., 2018), and allow multiple selection of files across simultaneously open folders (Sinha & Basu, 2012a).

**Alternative FM interaction paradigms.** Files and folders do not exist *in* the computer as literal, physical paper files and folders, of course, but are presented in this metaphorical way to provide users with a familiar idea of what digital objects are like and what can be done with them. The paper metaphor and the hierarchy provided with it are not the only possible way to represent and enable interaction with digital objects, and since the adoption of digital files many systems have been developed to implement alternative approaches (Burton, 1981; Burton, Russell, & Yerke, 1969). Some inherent limitations to non-hierarchical approaches like flat, linear, and spatial networks are discussed by (Indratmo & Vassileva, 2008). Below we review studies of systems implementing and testing these and additional approaches.

One theme among systems using non-standard approaches is metaphors that rely on common phenomena in human experience, such as space and time. This is achieved, for example, by putting the files into a different two-dimensional space, like a topic map (Yang, Zhou, Wang, & Lee, 2012), or a three-dimensional space where users can arrange and automatically re-arrange (Agarawala & Balakrishnan, 2006) their documents into piles (Mander, Salomon, & Wong, 1992) and other arrangements (Robertson et al., 1998) in the same way they may be in physical space. This utilises the spatial metaphor already popular in modern computing while avoiding the folder hierarchy, and enables highly personalised user-made reminding cues (Bondarenko & Janssen, 2005).

Research done in information visualisation on how to display hierarchies of various kinds in efficient and usable ways is also directly applicable to the display of the folder hierarchy, and in fact folder tree structures are often the specific cases used to demonstrate various general approaches (Turo & Johnson, 1992; W. Xu, Esteva, & Jain, 2010). Such work has typically consisted of designing a novel approach and comparing it to various baselines (Kobsa, 2004; Merčun & Žumer, 2013), and has generally focused on visualising especially large trees (Plaisant, Grosjean, & Bederson, 2003) using various two- and three-dimensional approaches. The most prototypical of these visualisations include treemaps (space-filling rectangles) (B. Johnson & Shneiderman, 1991), of which several variations exist (Stasko, Catrambone, Guzdial, & McDonald, 2000; Turo & Johnson, 1992), and animated 3d trees (Robertson, Mackinlay, & Card, 1991).

Files may also be presented chronologically, for example by allowing the user to specify a subset of documents based on some property (file access time or otherwise) and presenting them as a chronologically sorted, two-dimensional array (Fertig, Freeman, & Gelernter, 1996b; Freeman & Gelernter, 1996; Wideroos & Pekkola, 2007). Both spatial and chronological representations of files entail compromise: presenting time as locations in space (on the screen) mixes metaphors, while piles are unstructured containers that are functionally identical to a flat list of folders (Treglown, 2000).

Novel systems also represent digital items without metaphors, however, and typically do so simply as discrete items (whether called files or otherwise) in flat lists or tables sorted, arranged, or grouped by their properties, such as name, type, size, author, content, topic, and so on (Collins, 2007; Collins, Apted, & Kay, 2007; Collins et al., 2009; Dourish, Edwards, LaMarca, & Salisbury, 1999a; Dourish, Lamping, & Rodden, 1999; Dubey & Zhang, 2012; H.-F. Xu & Chen, 2011). By utilising items' properties beyond name and folder location, and the fact that users remember these additional properties (Gonçalves & Jorge, 2008a), new interactions are enabled: users may retrieve from their collection by recalling an item's narrative (Gonçalves & Jorge, 2006), or following a path of associations, for example from a user-remembered event to an email in which it is discussed and then to a document that was attached to the email (Kim, Croft, Smith, & Bakalov, 2011). Classifying by property also allows users to assign items to multiple groups, rather than a single folder (Quan, Bakshi, Huynh, & Karger, 2003), thus avoiding the single classification problem of the folder hierarchy. Items' properties can then be used to present items according to a logical division of content, such as in easily understood Venn diagrams (De Chiara, Erra, & Scarano, 2003), or in robust relational databases (Marsden & Cairns, 2003). One broad possibility enabled by focusing on item properties has been to unify digital items of all types (emails, files, Web documents, etc.) and present them together (Völkel & Haller, 2009), for example grouped by their properties (Dong, 2005; Dumais et al., 2016); if effective in its execution, an integrated presentation of files and documents across local storage and the Web would help to alleviate issues of information fragmentation (Bergman, Beyth-Marom, & Nachmias, 2006; Capra et al., 2014) and provide a flexibility that more closely resembles the physical world (Bondarenko & Janssen, 2005).

Another approach is to utilise properties of the user, rather than properties of the digital items, and for this user activity, task, and context have been the most popular thus far. With this approach, digital items need not be categorised in the folder hierarchy, but instead can be presented in a two-dimensional space in clusters representing their relevant activity or task (Krishnan & Jones, 2005; Wideroos & Pekkola, 2006). This can be taken even further by providing computing environments and workspaces dedicated to specific work- and PIM-related activities (Jeuris, Houben, & Bardram, 2014), where only relevant programs are displayed, and suspended while changing tasks. However, demarcating a single task or activity is challenging; approaches to this include allowing users to generate activity names and apply them to files with tags (S. Voida & Mynatt, 2009; S. Voida, Mynatt, & Edwards, 2008), determining an activity by analysing the times when files are in use (Krishnan & Jones, 2005), and logging the instances and times of common software interactions (Chernov, Demartini, Herder, Kopycki, & Nejdl, 2008).

As discussed above, tagging has been investigated for its potential use in providing multiple classification of files, thus obviating maintaining a folder hierarchy. Several systems have implemented this, either by using tags without the folder hierarchy (Seltzer & Murphy, 2009) or in tandem with it (Albadri, Watson, & Dekeyser, 2016; Voit, Andrews, & Slany, 2011). The ubiquity of the tagging concept means it can be offered as an unobtrusive feature (Oleksik et al., 2009) in both local and Web-based FM systems (Hsieh, Chen, Lin, & Sun, 2008) and in document management systems (Ma & Wiedenbeck, 2009).

Most of these novel approaches have had little effect on file management beyond their initial testing. A tagging feature has been introduced to Mac's Finder application, however, where it is offered alongside the folder hierarchy. This may be the most drastic change that file management will encounter in the near future; because current operating systems deal with files, any software that aims to replace them must still provide users with some access to them (Kaptelinin, 2003), thus prolonging the habit of managing them and therefore the need for such functionality.

A table summarising the system augmentations, alternative FM metaphors, and related hierarchy visualisation studies is presented in Table 4.

## Theory and methodology in file management research

In this section we discuss how FM research is carried out, first by noting the current theoretical underpinnings adopted, second by examining the methods used to study user behaviour, and third by examining how systems and services are compared and improved.

| System augmentation | Studies |
| --- | --- |
| improved and assisted search (12) | Adrian, Klinkigt, Maus, and Dengel (2009), Chirita, Costache, Nejdl, and Paiu (2006), B. Cole (2005), Ghorashi and Jensen (2012), Handschuh, Möller, and Groza (2007), Kim and Croft (2010), G. Liu, Jiang, and Feng (2017), W. Liu, Rioul, McGrenere, Mackay, and Beaudouin-Lafon (2018), Manber (1994), Sauermann, Bernardi, and Dengel (2005), Sauermann et al. (2006), Seenu, Rao, and Padma (2014) |
| improved cloud and file sharing (7) | Bergman, Whittaker, and Frishman (2018), W. Jones, Thorsteinson, Thepvongsa, and Garrett (2016), C. C. Marshall, Wobber, Ramasubramanian, and Terry (2012), Prinz and Zaman (2005), Rode et al. (2006), S. Voida, Edwards, Newman, Grinter, and Ducheneaut (2006), Whalen, Toms, and Blustein (2008) |
| enriched file or folder metadata (6) | He, Li, and Shen (2013), W. Jones, Hou, Sethanandha, Bi, and Gemmell (2010), W. Jones, Thorsteinson, Thepvongsa, and Garrett (2016), Roy, Singh, Chawla, Saxena, and Sinha (2017), Tayeh, Ebrahimi, and Signer (2018), S. Voida and Greenberg (2009) |
| improved navigation (3) | Fitchett, Cockburn, and Gutwin (2013, 2014), Vicente and Williges (1988) |
| file or folder de-emphasis (3) | Bergman, Elyada, Dvir, Vaitzman, and Ami (2015), Bergman, Tucker, Beyth-Marom, Cutrell, and Whittaker (2009), Lee and Bederson (2003) |
| improved or assisted selecting, moving, copying (3) | Ishizawa, Andoh, and Nishida (2010), Sinha and Basu (2012a, 2012c) |
| assisted filing (2) | Prinz and Zaman (2005), Sinha and Basu (2012b) |
| assisted backup (1) | Cox, Murray, and Noble (2002) |

| Alternative metaphor | Examples |
| --- | --- |
| according to items' properties (19) | Collins (2007), Collins, Apted, and Kay (2007), Collins et al. (2009), Dourish et al. (2000), Dourish, Edwards, LaMarca, and Salisbury (1999a, 1999b), Dubey and Zhang (2012), Gifford, Jouvelot, Sheldon, and O'Toole (1991), Gonçalves and Jorge (2006), Haller and Abecker (2010), Hardy and Schwartz (1993), Kim, Croft, Smith, and Bakalov (2011), Mosweunyane, Carr, and Gibbins (2011), Quan, Bakshi, Huynh, and Karger (2003), Sajedi, Afzali, and Zabardast (2012), Salmon (2009), Schaffer and Greenberg (1993), Thai, Handschuh, and Decker (2008), H.-F. Xu and Chen (2011) |
| using tags (8) | Adrian, Sauermann, and Roth-Berghofer (2007), Albadri, Watson, and Dekeyser (2016), Bloehdorn and Völkel (2006), Hsieh, Chen, Lin, and Sun (2008), Oleksik et al. (2009), Seltzer and Murphy (2009), Voit, Andrews, and Slany (2011, 2012b) |
| spatially (7) | Agarawala and Balakrishnan (2006), Altom, Buher, Downey, and Faiola (2004), Bauer, Fastrez, and Hollan (2005), Mander, Salomon, and Wong (1992), Robertson et al. (1998), Sinha and Basu (2012b), Yang, Zhou, Wang, and Lee (2012) |
| by relevant activity (6) | Dragunov et al. (2005), Hirakawa, Mizumoto, Yoshitaka, and Ichikawa (1998), Jeuris, Houben, and Bardram (2014), Shneiderman and Plaisant (1994), S. Voida and Mynatt (2009), Wideroos and Pekkola (2006) |
| chronologically (4) | Fertig, Freeman, and Gelernter (1996b), Freeman and Gelernter (1996), Gyllstrom (2009), Wideroos and Pekkola (2007) |
| logically (3) | Bowman, Dharap, Baruah, Camargo, and Potti (1994), De Chiara, Erra, and Scarano (2003), Marsden and Cairns (2003) |
| integrated, combining approaches (9) | Cutrell (2006), Cutrell, Robbins, Dumais, and Sarin (2006), Dittrich and Salles (2006), Dong (2005), Dumais et al. (2016), Gemmell, Bell, Lueder, Drucker, and Wong (2002), Krishnan and Jones (2005), Nelson (1999), Völkel and Haller (2009) |

| Related information visualisation studies | Examples |
| --- | --- |
| visualising hierarchies | B. Johnson and Shneiderman (1991), Kobsa (2004), Merčun and Žumer (2013), Plaisant, Grosjean, and Bederson (2003), Robertson, Mackinlay, and Card (1991), Stasko, Catrambone, Guzdial, and McDonald (2000), Turo and Johnson (1992), W. Xu, Esteva, and Jain (2010) |

**Table 4**

*Studies exploring FM software augmentation or alternative metaphors to files in a folder hierarchy, and studies focusing on related problems in hierarchy visualisation.*

**Theoretical and conceptual frameworks**

There do not currently exist any explicit theories about file management or theoretical frameworks specifically for understanding it. Similarly, to our knowledge no philosophical positions have been discussed in relation to FM or PIM, and the predominant implicit positions taken in FM research are post-positivist or constructivist. Put very briefly, post-positivism takes human perceptions and scientific measurements to be of a *real* world where causes reliably determine effects, but acknowledges the impact of various biases in our knowledge about that world and the methods we use to acquire such knowledge. This position is generally assumed when using quantitative approaches to scientific inquiry. By contrast, a constructivist position, which holds that the world is *constructed* by and consists only of perceptions and interpretations, is typical of qualitative approaches seeking to identify how meaning and behaviour are constructed and conceived of (Bryman, 2012). Both approaches may be useful in FM depending on the research questions being asked, as may the many positions in the spectrum between the two, but careful consideration of how these influence the questions, methods, and conclusions of FM research has yet to be carried out.

There are, however, two models for broadly characterising user behaviour in PIM, and these account for and thus apply straightforwardly to FM contexts. Each characterises user behaviour as describable according to one of three categories; for W. Jones (2007a), these categories are keeping, finding or refinding, and organising (also called metalevel) behaviour, while for Whittaker (2011) these are keeping, exploiting, and managing. The two models are similar, and since exploiting or utilising information often entails (re)finding it, those categories could be collapsed into one (e.g., refinding and utilising), making the approaches roughly equivalent. Alternatively, exploiting and refinding could be kept distinct and serialised (e.g., one refinds and *then* utilises information), making the models complementary. Regardless, these models capture the main concerns of traditional PIM and FM, as is reflected in the sections above that summarise user behaviour. Currently missing from each model, however, is explicit mention of the increasingly social aspect of personal information, which consists not only of co-managing (captured by metalevel or managing) but also of sharing (i.e., sending, receiving, and so on) information.

The conceptual frameworks and as-of-yet inactive theoretical landscape of FM are summarised in Table 5.

**Methods for understanding user behaviour**

Three general approaches to studying FM behaviour can be identified in the literature, and are often used together: ask participants about their behaviour, observe the

| Concept | Summary |
|---|---|
| Theories, philosophical positions | *There has not yet been discussion of theory or philosophical positions as they relate to file management research. Philosophical positions are generally implicit, and either post-positivist in quantitative studies or constructivist in qualitative ones.* |

| Author | Conceptual framework of PIM |
|---|---|
| W. Jones (2007b) | keeping, (re)finding, and managing (metalevel) information |
| Whittaker (2011) | keeping, exploiting, managing information |

**Table 5**

*Conceptual and theoretical frameworks that have been discussed for PIM and are applicable to FM.*

behaviour directly, and infer the behaviour from the file system. We examine each in turn.

**Asking.**  Asking participants about their file management behaviour has typically been done to discover user behaviour and challenges and understand the relevant contexts, usually by capturing participants' responses with digital questionnaires or recorded interviews. For example, studies using this approach have examined the challenge of coordinating files across multiple devices (Capra, 2009; Song & Ling, 2011), difficulties in managing files in Mac OS (Ravasio et al., 2004), students' habits in downloading documents (Huvila et al., 2014), opinions about graphical file management (Kaptelinin, 1996), and user perceptions about searching for files (Bergman, Beyth-Marom, Nachmias, et al., 2008; Teevan et al., 2004). It is rarely the only approach used in a study; rather, it is combined with the approaches described below when user perceptions are needed to understand the observed or inferred behaviour ((Whitham & Cruickshank, 2017) for example, combine all three approaches).

This approach is direct, as data about user perceptions and behaviour can be gleaned from participants rather than inferred from their behaviour. As with other forms of ethnographic study the data collected can be rich and useful for understanding contextual factors and informing the design of relevant systems and services. One disadvantage, however, is that users may not have accurate knowledge about their own behaviour: one study found a large discrepancy between users' attitudes about tagging their files (e.g., very positive) and their actual tagging behaviour (e.g., they typically did not tag files even when a good tagging system was presented and explained to them) (Bergman, Gradovitch, Bar-Ilan, & Beyth-Marom, 2013). Further, they may simply not be aware of any number of details about their own behaviour; for example, it is unlikely that anyone is cognisant of the number of redundant files they keep.

**Observing.**  Observation is a popular approach to investigating fine-grained phenomena and specific challenges in FM. Studies using this approach have, for example, sought to understand if digital documents are organised like paper documents (Barreau, 1995), how information from the Web is stored in files (W. Jones et al., 2001; W. Jones et al., 2002), and various challenges of file retrieval (Bergman et al., 2014). This is typically done by recording participants as they perform ordinary work, prompted retrieval tasks (e.g., elicited personal information retrieval or EPIR tasks as in Bergman et al., 2018, among others), or a guided tour (Thomson, 2015), where they navigate and explain their folder arrangement to an observer and perform common file management tasks along the way. Observation notes stored on paper, video recordings, and screen shots are all relatively straightforward methods that have been used to capture data, although unclear recordings have resulted in lost data (Bergman et al., 2010). Complex methods for observing users in more fine-grained ways have also recently been used (Benn et al., 2015).

Using this approach, actions are observed as they occur semi-naturally (e.g., during work) or when solicited (e.g., in a structured task or guided tour). Observation of this sort always takes place during some time, however, and thus necessarily does not see what participants are doing when not observed. This may be alleviated by supplementing observations with logs and inferences drawn as discussed next; for example, file creation, access, and modify times stored by the operating system can provide evidence of what participants do between immediate observations.

**Inferring.**  Users' actions determine properties of their file systems and the files and folders; for example, the folder hierarchy depth, the types of files stored, and the size of the collection in bytes and in total files and folders are each the result of specific user actions to store and organise their digital items, and their properties provide traces of this behaviour. The file system therefore serves as an artefact from which we may infer users' past behaviour, and studies have used this to study FM since the 1980s. They have, for example, observed files' sizes (A. J. Smith, 1981) and names (Carroll, 1982), examined how files are organised into folders (Khoo et al., 2007), explored the role of provenance in file retrieval (Jensen et al., 2010), studied the document management behaviour of students (Henderson & Srinivasan, 2011), and examined the effect of folder depth on file retrieval (Bergman, Whittaker, Sanderson, et al., 2012).

This approach has been implemented in two ways, which are used approximately as frequently and sometimes together. First, researchers have examined the file system as it appeared in the recordings of the interviews, guided tours, or structured tasks described above. This method is accessible as it does not require developing software, but given the large number of observable file system properties discussed below, manual notation of the properties is necessarily either highly laborious to collect and analyse or else limited to a small number, and it does not allow for capturing properties of portions of the file system not seen during the task or tour (Bergman et al., 2010). A second way is to use custom software to traverse the folder tree, recording data about the files and folders encountered, or to log user actions or changes to files and folders.

Automated methods facilitate studying a large sample and many variables (e.g., file system properties), including temporal data, but are a technical challenge to develop and implement (Dinneen, Odoni, Frissen, & Julien, 2016). Both manual and automatic collection methods require participant trust to let the researcher, possibly perceived as an expert in PIM, see their digital organisation or perceived lack thereof (Barreau, 1995). Automatic methods may also require researcher supervision to use, thus restricting sample size by being difficult to administer, or may be an obstacle to recruitment because it is difficult to find users willing to expose and share their digital possessions and desktops, entailing that participants are from an available but niche group like trusting colleagues. It is also difficult to develop such software to support multiple operating systems; perhaps as a result, researchers have instead relied on tools packaged with the OS (e.g., Evans & Kuenning, 2002) that provide minimal functionality, and typically focused on a single OS (e.g., Khoo et al., 2007).

A look at thirty-one studies examining the file system reveals the use of these methods, the number of participants in the sample, and the file system properties examined (presented together in Table 6). It should be noted that a low number of file system properties or a small sample size does not necessarily indicate an ineffective FM study or researcher oversight, as studies have explored differing research questions requiring collecting data about only particular file system properties.

Twenty-eight properties of the file system have been examined across the studies mentioned above, regardless of the data collection method used. This includes five variables that are particularly important to general PIM contexts: collection size, folder depth, folder breadth, folder size, and redundancy (e.g., in file and folder names) (Bergman, 2013). Twelve additional properties were suggested by Dinneen, Odoni, Frissen, and Julien (2016), resulting in forty properties available for use in FM research (presented in Table 7).

Together, these properties characterise each category of FM behaviour discussed above, and in smaller groups provide insight into particular actions and challenges users regularly encounter. The most commonly made measurements include folder tree height, breadth, number of subfolders per folder (sometimes called *branching factor*), and consistency (usually defined as deviation of branches from the average), which inform us, respectively, of the maximum depth to which users may need to traverse to find a file, the maximum (breadth) and average (branching factor) number of folders competing for a user's attention at any depth, and the likelihood of the user encountering an unfamiliarly structured area (or branch) of the tree during navigation. The *time of last access* of files

| Study | n = | Data collection methods | FM properties examined |
|---|---|---|---|
| Satyanarayanan (1981) | 8 | simple software | collection size; file size |
| Carroll (1982) | 22 | structured task | file type; collection size; file name; length of name |
| Akin, Baykan, and Rao (1987) | 171 | structured task | branching factor; folder fullness; folder depth; file and folder names |
| Bennett, Bauer, and Kinchlea (1991) | 3 | simple software | collection size; file size; use of symbolic links; file types; number of folders |
| Sienknecht, Friedrich, Martinka, and Friedenbach (1994) | 267 | simple software | file size; collection size; files per folder; branching factor; file access |
| Barreau (1995) | 7 | guided tour | file names; file access times; use of default locations |
| Nardi, Anderson, and Erickson (1995) | 15 | guided tour | file type (ephemeral, working, or archive) |
| Douceur and Bolosky (1999) | 10,568 | simple software | file size; files per folder; folder depth; file creation and modification; file types; leaf folders |
| Vogels (1999) | 45 | simple software | file size; file type; collection size |
| Downey (2001) | 562 | simple software | file size |
| Evans and Kuenning (2002) | 22 | simple software | file type; file size |
| Gonçalves and Jorge (2003) | 11 | simple software, interview | tree depth; total file count; branching factor; files per folder; file types; file size; file creation, modified, accessed times; use of numbers, whitespace, and punctuation in names; length of file names; use of shortcuts/symlinks |
| Boardman and Sasse (2004) | 31 | simple software, guided tour, diary | total folders; folder depth; unfiled files |
| Ravasio, Schär, and Krueger (2004) | 16 | guided tour | file age; files per folder; use of desktop |
| Henderson (2005) | 6 | simple software, interview | total folders; file names; duplicate file names; duplicate folder names; branch consistency |
| W. Jones, Phuwanartnurak, Gill, and Bruce (2005) | 14 | guided tour | branching factor; file types; file names |
| Agrawal, Bolosky, Douceur, and Lorch (2007) | 62,744 | simple software | file size; collection size; file types; file creation and modification; files per folder; use of default locations; file depth; folder count |
| Khoo et al. (2007) | 12 | simple software, interview | use of default folders; roots per user; use of desktop; tree height and breadth; files per folder; file names |
| Hicks, Dong, Palmer, and McAlpine (2008) | 40 | simple software, questionnaire | file names; tree depth; file depth; file size; collection size in bytes; file types; file and folder duplication (by name); file access times |
| Hardof-Jaffe, Hershkovitz, Abu-Kishk, Bergman, and Nachmias (2009b) | 518 | custom online environment | collection size; tree dimensions; files per folder; file depth; unfiled files |
| Henderson and Srinivasan (2009) | 73 | simple software | collection size; tree height; file depth; branching factors; root folders; file name duplication, folder name duplication; empty folders |
| Bergman, Whittaker, Sanderson, Nachmias, and Ramamoorthy (2010) | 296 | structured task | file depth; use of desktop; use of shortcuts; files per folder; branching factor; use of default locations; files per folder |
| Henderson (2011) | 73 | interview | unfiled files; tree height; file name duplication; folder name duplication; use of desktop; use of default locations |
| Henderson and Srinivasan (2011) | 10 | interview, simple software | unfiled files; tree height; file name duplication; folder name duplication; use of desktop; use of default locations |
| Bergman, Whittaker, Sanderson, Nachmias, and Ramamoorthy (2012) | 289 | structured task | file depth; files per folder; branching factor |
| Bergman, Whittaker, and Falk (2014) | 275 | structured task | file depth; file type; file access time |
| Massey, TenBrook, Tatum, and Whittaker (2014) | 62 | simple software | total files; use of desktop; file types |
| H. Zhang and Hu (2014) | 12 | guided tour, simple software | tree breadth, tree shape, files per folder, branching factor, total files, folder depth |
| Fitchett and Cockburn (2015) | 26 | interview, logging | file access; file types; file name length; file depth; use of desktop |
| Benn et al. (2015) | 17 | structured task | folder depth |
| Whitham and Cruickshank (2017) | 12 | guided tour, scan and logging software | total files, total folders; file and folder access and modify times; |

## Table 6

*Studies observing participants' file systems, number of participants, data collection method, and file system measures reported. 'Simple' in this case is not evaluative but rather distinguishes, for example, data collection scripts from persistent logging software.*

and their depth in the folder hierarchy can help to quantitatively describe users' archiving habits, and the number of duplicated file and folder names indicate the difficulty they face in differentiating and naming similar items in their collections (Henderson, 2011). Properties can also be examined for correlation with individual difference and external factors, for example to see if certain occupations or personality styles correlate with the average length of file names or total number of files (Massey, TenBrook, et al., 2014). The varied goals and research questions present across studies of this type entail that despite collectively looking at many of these properties, a complete quantitative description of general FM behaviour (i.e., storage, organisation, retrieval...) does not yet exist and cannot be derived from cross-study analyses (e.g., meta-analyses). The implications of and suggested solution to this are discussed in the future research directions, below.

| FM topic | Data about | Properties |
|---|---|---|
| Storage | Hardware (4) | # of available drives, hard drive capacity, use, and free space; |
| | Collection (13) | total files, total folders; collection size (in bytes), collection size (files + folders); file extensions/types; file sizes; file age, folder age; shortcuts/symlinks, hidden files, hidden folders; duplicate files (by hard link), duplicate folders (by hard link) |
| | Semantics (7) | File or folder name, length of name, numbers in names, punctuation or special characters in names, duplication of names; Letters in names, whitespace in names |
| organisation | Structure (12) | Root folders; tree breadth, tree depth; folders in each folder (branching factor), files in each folder; file depths, folder depths; branch consistency or skewness; use of desktop for storage, use of default folders; inaccessible folders in user space; folders excluded by participants from study |
| Retrieval | File access (4) | File access times, file modify times; folder access times, folder modify times |

**Table 7**

*40 properties of file system (e.g., as measured to study personal digital collections and infer users' FM behaviour).*

## Methods for designing and evaluating FM systems

As noted above, the general process for improving existing and novel FM systems and approaches entails evaluating systems' performance or users' performance or preference, for example during structured tasks and in comparison to some baseline system. However, it is agreed among researchers that meaningfully evaluating and comparing PIM systems is extremely challenging (Kelly, 2006), due to four factors that apply as much to FM as they do to broader PIM contexts. First, PIM behaviour is complex and idiosyncratic, so the relative effects of the many factors can be difficult to understand and it is not always clear which tasks are best for an experiment (Capra & Perez-Quinones, 2006; Kelly, 2006). This is compounded when performing longitudinal studies, as user behaviour across time is not well understood; longitudinal approaches to evaluating FM are thus rare (Dinneen, Odoni, & Julien, 2016). Second, representative data sets (i.e., test collections) do not currently exist, preventing apples-to-apples comparisons in evaluating FM systems (e.g., system and user performance), and one may reasonably doubt the possibility (or usefulness) of generalised FM collections – and also of generalised PIM models – given that PIM is, by definition, such a highly personalised domain. Such efficacy may soon be empirically tested, however, as the possibility of creating such collections and models draws near: recent methodological contributions have enabled observing extensive file system properties across many participants (Dinneen, Odoni, Frissen, & Julien, 2016), thus constituting a step towards generating representative and generalised test collections, and activity logging may be used to model user behaviour (Chernov et al., 2008). Third, traditional evaluation measures do not apply straightforwardly to FM contexts; for example, recall and precision are of limited use in FM retrieval evaluation, as most FM retrievals are looking for a particular file rather than a large batch of files (Fitchett & Cockburn, 2015), and it is impractical to ask a single participant to make relevancy judgements for all of their documents and invalid to ask third parties to help in this (Gonçalves & Jorge, 2008b). Fourth, though it is essential for carrying out valid comparative evaluations, it can be difficult to make fair comparisons between systems and approaches when they are created with differing affordances and intended interactions (Voit, Andrews, & Slany, 2012a). One approach to comparing efficacy, efficiency, and usability across disparate systems is by doing an evaluation called GOMS model analysis (Kieras, 1997), which can provide an outcome-based comparison in cases where possible user behaviour can be enumerated and predicted with some confidence. This has been used, for example, for testing the efficiency in moving and deleting files in a new file manager as compared with the existing File Explorer (Sinha & Basu, 2012a).

Evaluation aside, an explicit approach to the general design of PIM systems that clearly applies to FM systems is the *user-subjective approach* (Bergman, Beyth-Marom, & Nachmias, 2003), which emphasises that PIM tool design should be concerned with what the users find important, rather than studying only how users behave with current, limited systems. This approach has been explicitly utilised in several studies (Bergman, 2012; Bergman, Beyth-Marom, & Nachmias, 2008; Bergman et al., 2009), and so shows promise for FM-specific software design. Examples of literature pertaining to reflection on the design and evaluation of FM systems are presented in Table 8.

| Topic | Examples |
| --- | --- |
| System design | Bergman (2012), Bergman, Beyth-Marom, and Nachmias (2003, 2008) |
| Experiment, task, data set design | Capra and Perez-Quinones (2006), Chernov, Demartini, Herder, Kopycki, and Nejdl (2008), Dinneen, Odoni, Frissen, and Julien (2016), Dinneen, Odoni, and Julien (2016), Gonçalves and Jorge (2008b), Kelly (2006), Voit, Andrews, and Slany (2012a) |

**Table 8**
*Examples of literature relevant to the design and evaluation of FM systems.*

## Discussion

We discuss here the importance and relation of FM research to various research areas, and then discuss the future directions and challenges facing FM research.

### Importance to other research areas

By virtue of studying how humans use computers to manage information, FM research shares the concerns and methods of research areas such as personal information management, computer supported collaborative work, information retrieval, and human-computer interaction. FM also has broad import for core subfields in information sciences, like information behaviour, information organisation, and personal archiving, and closely related fields like management information systems. Finally, its greater context and wide range of possible determining factors entail that it even has overlap and potential implications for psychology, computer science, and philosophy. To further explicate the importance of FM and begin suggesting future research directions, we discuss such overlaps in this section. Some particular topics within the overlaps have been suggested in past work, and some are newly proposed here, but most remain relatively unexamined; we hope discussing these topics here spurs readers to undertake work on them, thus realising potential value in FM research.

**Personal information management.** The research area most closely related to FM research is personal information management (PIM), which can be argued to be the broader *parent* topic to which FM research belongs, although this has not yet been explicitly posited and defended. For example, in this view FM can be seen as a subset of PIM focusing specifically on how people manage information at the file and folder level. The contexts of files and folders is arguably of crucial importance in PIM, given that much of the information of our daily lives resides in the digital domain, specifically in files. Indeed, the categories of research described above could be used to describe common concerns in PIM: to understand peoples' behaviour when personally managing information, to understand what gives rise to differences in this behaviour, and to improve the design of the relevant systems and services that support this. Typical FM activity accords with the various conceptions of *personal* fundamental to and used in PIM literature; for example, that *personal* includes being controlled by, owned by, about, directed toward, sent by, experienced by, or potentially relevant to an individual (W. Jones et al., 2017). Unsurprisingly, files and folders have been present in and are relevant to many PIM studies that focus on the management of digital items by type or format, including digital music collections (Brinegar & Capra, 2010, 2011), digital photo collections (Rodden & Wood, 2003), and scholarly references (Fastrez & Jacques, 2015). Though users have the option to manage these digital items within their respective format-specific applications, they may also manage them as files, and insights gleaned from FM studies have implications for their general management. More about these two modes of management, of digital items as files or as specific formats, is discussed below.

**Human-computer interaction, information retrieval, computer-supported cooperative work.** FM research has relevance to human-computer interaction (HCI), information retrieval (IR), and computer-supported cooperative (or collaborative) work (CSCW), and this is reflected in the presence of PIM workshops in the last decade at the relevant HCI (2008 at SIGCHI, 2016 at CHI), IR (SIGIR 2006), and CSCW (2012) workshops. Managing files is a required activity for anything beyond the simplest computer usage. Due to this ubiquity and fundamental nature, it is of considerable relevance to research in HCI, where the file-folder metaphor has been a common example of typical user interactions, for example in debates about digital design and manipulation philosophies (Frohlich, 1997) and the broader desktop metaphor (J. Johnson, 1987; Ravasio & Tscherter, 2007). It is within the HCI community primarily that the debate about the use of the file and folder metaphor, summarised above, has taken place. FM may also serve as an excellent context for advancing our knowledge of information foraging theory (Pirolli, 2007), which is of interest to those studying HCI and information behaviour (IB) alike; with folders and files serving as metaphorical bushes and berries, it is reasonable to describe users' FM behaviour as enriching their file systems by storing and organising, following scents by navigating, and foraging by retrieving.

Because much FM activity consists of retrieval or is done to support later retrieval, it is reasonable that FM research also has a close connection with IR research. The role of search (both for files and through files) in FM has been a focal point of FM research, and this has provided insights into how users retrieve files with search, navigation, or both, as discussed above. FM systems and their users benefit from innovations in IR research, for example in the retrieval and ranking algorithms and improved full-text and faceted search. Similarly, FM research may benefit from methods used in IR research; for example, by analysing desktop search logs to understand users' file search strategies (Jansen, Spink, & Saracevic, 2000). Because the findings and system augmentations in FM research may save users time when retrieving files, FM is also likely relevant to narrower, applied areas of HCI and IR – especially those dealing with time-sensitive tasks – such as browsing and retrieving medical documents (J. D. Baker, 2012).

FM is relevant also to research into CSCW and a topic within PIM known as group information management (GIM), as the opportunities, challenges, and implications of co-managing shared files, especially for collaborative work, are likely generalisable to broader contexts. For example, a study (Bergman et al., 2014) of the impact of shared files on retrieval success participates in and has implications for FM in understanding users' refinding behaviour, IR in supporting user behaviour with better file search algorithms, and CSCW in understanding how the shared files have supported shared tasks.

**Information behaviour and information seeking.**   FM research also has relevance to core areas in information science, such as IB and information-seeking behaviour (ISB), as is reflected by the presence of two PIM workshops at the ASIS&T annual meeting (in 2009 and 2013). IB research, understood as investigating "how people need, seek, manage, give, and use information" (Fisher, Erdelez, & McKechnie, 2005, p. xix), is clearly related to both PIM and FM, where users create, manage, and retrieve information stored in files, thus exhibiting particular patterns of IB. Thus unsurprisingly, typical IB patterns like filing, archiving, and organising collected information (Meho & Tibbo, 2003) match very closely what users do with files as described in the FM strategies previously characterised (Berlin et al., 1993; Boardman & Sasse, 2004; Malone, 1983). The role files play in greater IB and ISB patterns has been touched upon tangentially in many studies of PIM and ISB, but given the prevalence of files this should be investigated further; changes in ubiquitous and fundamental information software such as a file manager will likely affect the information behaviour of various groups.

**Personal archiving.**   File management research also has a promising but so far largely implicit overlap with work in personal archiving (PA) or personal digital archiving (PDA). One cause of this overlap is evident in the fact that:

> "what we have written, what we have read, where we have been, who has met with us, who has communicated with us, what we have purchased, and much else is recorded digitally in increasingly greater detail in personal digital archives, whether they are held by individuals, institutions, or commercial organisations, and whether we are aware of those archives or not" (Hawkins, 2013, p.2).

For those digital archives that are personal in virtue of being managed or owned by an individual person, FM is very likely taking place, and may be done either neglectfully, thus under-facilitating later reuse, or painstakingly, and could thus benefit from thoughtfully designed software support.

Numerous studies consider a person's files as being part of their personal archive or digital possessions collection (Cushing, 2013; Kaye et al., 2006; C. C. Marshall, Bly, & Brun-Cottan, 2006; C. C. Marshall, McCown, & Nelson, 2007; Massey, TenBrook, et al., 2014; Siddiqui & Turley, 2006), and some have begun to describe problems shared by FM and PA research alike, such as file ownership and disambiguation (Haun & Nürnberger, 2013). That some files are kept and preserved is certainly of interest to FM research, and it is clear that, say, file management augmentations could be designed specifically to support personal archiving. The potential for PIM and FM research to increase understanding of personal digital archiving has previously been suggested (Bass, 2013), and the overlap between the fields is being increasingly explored (e.g., at the Personal Digital Archiving 2017 conference, wherein PIM was a topic of discussion).

    **Knowledge organisation.**   FM research also has potential import for research in the organisation of information or knowledge, also called knowledge organisation (KO), which is concerned with the nature and quality of systems used to organise knowledge (usually in documents). Labelled folders and their parent-child relationships present users with a free-form way to structure and name information as they want to, and so identifying how and why they do so may produce insights for KO systems design in general. Identifying trends across adequate numbers of users would mean establishing reflections of current practices and expectations of document organisation tools (folder trees in this specific case), which should be considered when designing KO systems. For example, there is an open question in KO about how to best organise knowledge structures to aid interactions like retrieval, browsing, and exploration (Julien, Tirilly, Dinneen, & Guastavino, 2013); FM studies characterising people's folder trees (e.g., mean depth and breadth) tell us what kind of structures they are most accustomed to navigating, and thus suggest what shape KO hierarchies could take to leverage that familiarity. Conversely, works in KO have demonstrated the benefits to users of making dynamic changes to unfamiliar knowledge structures (Dinneen, Asadi, Frissen, Shu, & Julien, 2018), and so perhaps similar dynamic changes could be made in file-folder contexts (e.g., to facilitate new users' familiarisation with and use of shared drives).

    Research in KO has also been concerned with identifying highly skewed (e.g., power law) distributions (Smiraglia, 2002) in collections organised by groups of individuals; for example, the assignment of documents to Library of Congress subject headings is highly skewed, with most documents being assigned to a small number of subjects, and the many remaining (i.e., most) subjects therefore providing access to only a few documents each (Julien et al., 2013). Such distributions may be present in file systems as well (i.e., most folders may contain small numbers of files while a small group of folders contain most files), suggesting that structures created by an individual reflect those created by groups, or conversely, individually created structures may not be distributed like group-created structures, implying that group-made structures are unfamiliar in an additional way. Finally, the field of KO has begun to focus on individual differences relatively recently

(Rowley & Hartley, 2008), and so this too may be a valuable topic for which FM research could provide input (e.g., factors determining the creation and use of one's own folder structure may be relevant when browsing folder or subject heading structures made by others).

**Library services and cultural heritage institutions.** Many aspects of personal information management, including file management, and the resulting tasks, may manifest in the identification, use, and instruction of library services. Therefore, it is perhaps unsurprising that PIM and FM have explicitly concerned those looking to improve such services (Fourie, 2011, 2012; Otopah & Dadzie, 2013) even as long ago as 1989:

> "It is a natural extension of the librarian's increasingly computer-oriented information managing skills to act as a consultant to his or her institution concerning personal file management (PFM) software. PFM software automates and greatly improves a physician's or researcher's manual system of filing and managing a reprint collection. These personal collections of medical literature are often depended upon as a primary information source... It is natural for the librarian rather than for others, such as the institution's computer department staff, to teach PFM skills" (Strube, Antoniewicz, Glick, & Asu, 1989).

To our knowledge, PIM is not a part of the instruction librarians receive (e.g., in a master's of information studies), but the works discussed here suggest such instruction would be useful. This suggests work remains to identify relevant PIM (or PFM) skills, determine how to best convey them to patrons or clients in a position to benefit from them, and evaluate the benefit of doing so.

Similarly, FM and personal archives likely interest cultural heritage institutions (especially *digital* cultural heritage institutions), which are concerned with, among other things, the personal collections of relevant individuals and the ways in which those collections are organised and therefore preserved (e.g., original order). This, in turn, influences what value can later be made of such collections by libraries, archives and museums. The potential overlap of FM research with library services and cultural heritage indicates the relative importance of FM for the study of such fields, or library and information sciences (LIS). While the exact relationship of LIS to the more general information studies (IS) is beyond the scope of the present discussion, it could be argued that considering the relevance of FM to LIS, and personal digital archiving and information behaviour, as discussed above, FM is therefore of broad import in information studies or sciences (IS).

**Information systems.** FM research may also be of import to information systems research (i.e., management information systems), where special attention is paid to the use, adoption, implementation, and so on of IT, including software, in organisational contexts. Example topics likely of interest include the adoption or resistance to adoption of content management systems (M. L. Johnson et al., 2009) for securely sharing files, and decreasing onboarding time by simplifying shared drives (e.g., Dinneen et al., 2018). Conversely, influential information systems research examining the factors determining IT adoption, like the technology acceptance model and its recent extensions (Marangunić & Granić, 2015), may be useful in studying the adoption of file management software or cloud storage. The intersection of information systems and information behaviour has also received some attention (Johnstone, Bonner, & Tate, 2004), further suggesting the potential for IB, and consequently PIM and FM, in information systems research. It is perhaps unsurprising, then, that several of the works reviewed above have appeared in information systems venues like *ACM Transactions on Information Systems*.

**Computer science.**   It is also reasonable to infer a possible relevance of FM to computer science, where a considerable body of existing literature aims to understand the contents and access patterns of file systems, such as file size distribution (Tanenbaum, Herder, & Bos, 2006), to optimise hardware, firmware, and software. FM studies focusing on real-world file systems that users have interacted with may provide valuable data sets for such design goals, especially given that most of such computer science studies have examined only files stored on servers and software development machines.

**Psychology and philosophy.**   FM research and the file-folder paradigm may also be useful in fields beyond those concerned with the information and information systems. We have discussed above the psychological aspects of FM previously examined, but the relevance of FM to psychology may extend beyond this. For example, metaphors, metaphorical thinking, and categories and categorical thinking are common objects of study in psychology and prominent in FM (digital and analogue) and PIM generally (Case, 1991). It is not a stretch to think that other dimensions of individual difference are factors in FM, including those concerning psychology, like cognitive styles, and decision making processes (Kozhevnikov, 2007). At its broadest, general trends in file management studies may also be of interest to those studying topics like Philosophy of Information and Philosophy of Computing, which seek to understand what is possible in the digital realm, how much information we are storing as a society (Lyman & Varian, 2003), and to what extent humanity has moved into the infosphere (Floridi, 2010).

**Summary of importance to other research areas.**   FM has significance and potential import for several fields, including those within and nearby to information science. The problem – and opportunity – of having implicit and unrealised connections to nearby areas, is perhaps inherited from PIM research: over a decade ago William Jones commented on the unrealised synergies between PIM and other fields, pointing out in particular that works on information seeking (including sense-making) and information encountering have clear but unexplored import for PIM and vice versa (W. Jones, 2007b). Here we have identified further synergies between FM (and so PIM) and other topics; a crucial next step is to explore these potential synergies, both conceptually and in practice (e.g., by purposefully applying relevant information-seeking models to FM studies). Table 9 lists the fields and disciplines, discussed in this section, that have connections to FM research. We next discuss the future of FM and its study.

| Field or discipline related to FM | Abbreviation |
|---|---|
| Computer science | CS |
| Computer-supported cooperative work | CSCW |
| Group information management | GIM |
| Human-computer interaction | HCI |
| Information behaviour, information-seeking behaviour | IB, ISB |
| Information retrieval | IR |
| Information science or studies | IS |
| Library and information sciences | LIS |
| Organisation of information, knowledge organisation | OI/IO, KO |
| Personal archiving, personal digital archiving | PA, PDA |
| Personal information management | PIM |

**Table 9**
*List of fields and disciplines connected to file management research, with field abbreviations used in this paper.*

## Future challenges and research directions

In this section we present a discussion of the future challenges and directions in FM research that is structured to reflect the existing areas of research identified above, and have included at the end a discussion of the future of files and their management systems.

**Improved understanding of user behaviour.** Future research into users' behaviour will likely benefit from combining complementary insights from qualitative and quantitative studies, providing a complete picture of the various aspects, scope, and contexts of behaviour. Particularly, a broad quantitative description of the artifact produced by such behaviour (i.e., the file collection) would be useful for complementing the rich characterisations of users' FM behaviour that has emerged from the many qualitative descriptions, and may unify the disparate quantitative descriptions discussed above. This could comprise, for example, an extensive quantitative description of digital file collections' scale, structure, and contents. The results of such a description may then enable further methods for understanding user behaviour, like principal component analysis, user modelling, and the generation of a standardised, representative data set (i.e., test collection) for FM system evaluation (Chernov et al., 2008).

Such a quantitative picture cannot currently be produced by meta-analysis or collation of results, however. As noted above, one consequence of the varying goals and research questions of previous studies is that many study contexts are fundamentally incomparable; for example, where one study examines the retrieval of recently used files seen during a controlled experiment (Bergman et al., 2010), another examines the folder structures created by students in an online learning environment during a class assignment (Hardof-Jaffe, Hershkovitz, Abu-Kishk, Bergman, & Nachmias, 2009a), and it is difficult to compare results across studies of public computers (Vogels, 1999) and servers (Sienknecht et al., 1994), or of only media files (Evans & Kuenning, 2002), shared files (Bergman et al., 2014), or recently accessed files (Fitchett & Cockburn, 2015). Another natural consequence of studies' varying goals is that even when contexts have been comparable, the measures collected and reported have typically differed; for example, studies of the file system have collectively looked at 28 of the 40 (or more) potential file system measures (Dinneen, Odoni, Frissen, & Julien, 2016), but typically with few measures per study (mean 4.4), and whereas one study reports (among other measures) the maximum depth at which folders are stored (Henderson & Srinivasan, 2011), another reports the average depth of currently used files (Fitchett & Cockburn, 2015).

While the description of digital collections outlined here cannot be derived from existing studies, it could be the explicit goal of future studies; specifically, future studies may examine as many of the available file system measures as possible, including those to help identify the scale, structure, and contents of collections, and in as generalisable of a context as possible. Such study will require data collection tools capable of overcoming or circumventing the limitations of current tools identified above. Once a more complete understanding of FM behaviour is achieved, fuelled by both qualitative and quantitative insights, it will be possible to compare aspects of FM user behaviour (e.g., the the structure of the collections created) to those in similar contexts, like the management of Web browser bookmarks (Kaye et al., 2006) or emails (Ducheneaut & Bellotti, 2001; Kalman & Ravid, 2015; Mackenzie, 2000).

Finally, time remains a challenge for understanding FM behaviour. Some of what is known about FM behaviour was established in studies that are now dated and possibly obsolete; for example, several (10) of the file system studies described above took place twenty years ago or more when the entire computing paradigm was drastically different: graphical interfaces were relatively new, storage was expensive, file name limits were much shorter, and perhaps more importantly, prior to the proliferation of Internet connections and individuals owning multiple computing devices, files were in greater isolation between machines. In other words, though the essential nature of file management has not changed, several aspects of it have; this is reflected in the now obscure media and terminology of older studies (e.g., tape storage; A. J. Smith, 1981). Beyond up-to-date studies, longitudinal studies will also help to address gaps in knowledge about how users' file collections and FM behaviour change over time. Although long-term management is a general concern of PIM (W. Jones, Bellotti, et al., 2016) and notable, longitudinal PIM studies reflect that concern (Boardman & Sasse, 2004; Čopič Pucihar et al., 2016; Dumais et al., 2016), longitudinal studies dedicated to FM have been scarce, perhaps because conducting such studies is a technical challenge (Dinneen, Odoni, & Julien, 2016).

**Improved understanding of determinant factors.**   From the above summary of research into the individual differences and external factors influencing FM behaviour one may reasonably conclude that further research is needed to understand and support for these factors. Factors like occupational traits, task and information type, the operating system, computer literacy, spatial ability, and personality style are not yet well understood, but may play significant roles in how users struggle or succeed in managing their files. Even the principled differences between the OSes in how users *can* manage files has not been made explicit. The default FM presentation style differs between the OSes, and this seems to affect the retrieval of recently used files (Bergman et al., 2010), but what of other system-based differences? Most of the details of these differences are scattered across user manuals and release notes, and have not been at the forefront of FM research despite their obvious influence.

The effects of individual differences on FM behaviour are also good candidates for future FM research, as no specific difference is well understood. For example, the two previous studies of spatial ability suggest that file management is influenced by general spatial cognition (Vicente et al., 1987; Vicente & Williges, 1988), but it is unclear if this extends beyond folder navigation (e.g., to folder organisation) and to what extent spatial ability specifically is responsible for such influence. The full relationship between personality and file management also remains unclear, as discussed above, and further additional individual differences are also likely playing determining roles in FM (e.g., personal need for structure; Neuberg & Newsom, 1993).

One particularly notable individual difference that may affect FM, but has yet to receive research interest in that context, is cognitive style, or the general way people think about information (Sternberg & Sternberg, 2017). Cognitive style has been studied for how it affects learning (Tsianos, Germanakos, Lekkas, Mourlas, & Samaras, 2009), decision making (Kozhevnikov, 2007), information seeking behaviour (Ford, Wilson, Foster, Ellis, & Spink, 2002), Web browsing (Chen & Rada, 1996), and Web search behaviour (Hariri,

Asadi, & Mansourian, 2014; Kinley, Tjondronegoro, Partridge, & Edwards, 2014). It is reasonable to infer that FM behaviour may be influenced by cognitive style, and in particular Riding's view of cognitive style, which integrates several views (Riding & Cheema, 1991), may be useful for examining this; it defines cognitive style as a preference for verbal- or image-based and analytic or wholistic information and thinking (Riding, 1997). This seems well-suited to studying FM, where users have opportunities to act on these styles and producing file and folder arrangements that reflect their style, for example by categorising files with many folders or synthesising them into a few, or by relying on folder names or images for retrieval.

**Improved systems and services.**   Applying the findings of previous studies to improved systems is a fertile area for future FM research. One direction for this is in helping users understand, whether analytically through information literacy or intuitively through system transparency, the FM metaphor and FM system capabilities. Users often do not understand files, digital content, the actions that can be done with a file, when those actions are appropriate or reliable, or who owns and can access a file (Brostoff et al., 2005; Harper et al., 2013; Odom, Zimmerman, & Forlizzi, 2011). Future systems should therefore not only be faster, enabling greater productivity, but also simpler, either enabling more accurate and easier mental modeling or precluding the need for it. This, in turn, requires continuing to identify specific confusions.

The development of future FM systems may be guided by existing considerations and opinions, for example that systems should improve upon existing systems by facilitating flexible, familiar, *ad hoc* restructuring (Bondarenko & Janssen, 2005; Indratmo & Vassileva, 2008; W. Jones et al., 2005), further unify personal information items from various sources (Warren, 2014), and act as a prosthesis for human memory to support intuitive and natural interaction (Trullemans & Signer, 2014b). The usability of FM software has not been previously touched upon, and is thus a promising direction for improving FM systems. This may be achieved, for example using a GOMS model (Goals, Operators, Methods, Selection rules) or by designing FM software for specific uses or user groups, such as new and casual computer users (Sinha & Basu, 2012a). Further, the design of such systems may benefit from being more deliberately and explicitly *design-like*, for example by applying the methods and iterative design cycles of design science research (e.g., as outlined by Peffers, Tuunanen, Rothenberger, & Chatterjee, 2007), which has, to our knowledge, not been explicitly done in FM or PIM research.

Though it has been identified as mainly supporting navigation rather than replacing it, search will likely continue to be an important research area. There are many potential improvements to be made to search, for example by improving the display and interactivity of file search results (G. Smith et al., 2006), further integrating search with navigation by using queries to guide navigation (Fitchett et al., 2014), or further still, creating two-way interactions between the file tree and search results, as has been done with LCSH (Julien et al., 2016). The evaluation of desktop search, where recall and precision are imperfect measures for reasons discussed above, may find benefit in the application of alternative measures, like mean reciprocal rank.

Building systems to support GIM and the social aspects of FM is promising. Currently, Dropbox and such software allows for synchronisation of individual file spaces, but as discussed above users often misunderstand where exactly these files are and what can be done with them. Something like the Dogear social bookmarking system (Millen, Yang, Whittaker, & Feinberg, 2007), but with successful integration of files, would likely be valuable in supporting users in tasks requiring collaborative FM. Views, or on-the-fly, ephemeral display of sets of folders and files, may help with this and with overcoming problems of mutual intelligibility (Dourish, Lamping, & Rodden, 1999), especially if unfamiliar folder structures are modified with hierarchy pruning algorithms (e.g., Dinneen et al., 2018; Julien et al., 2013), but it remains to be verified if such systems would empower or confuse users.

Designing and improving services, such as library services (Fourie, 2011), to support PIM and FM is a promising but difficult future research direction. Though information literacy and education initiatives may be designed to include FM and other aspects of PIM, it is first necessary to identify best practices so that recommendations can be made. That few prescriptions are derived from PIM research is surprising given the vast array of strategies for categorising and filing paper records and documents that were present and promoted in the 1980s (Gill, 1988) and the subsequent proliferation of the digital computer file; some individuals now have more files in the digital domain than organisations once had in paper, but fewer resources for organising them. These office file management strategies of the past may serve as inspiration for future digital organisation strategies, as they could accommodate a wide range of organisational approaches, but would need to be updated to account for the current, digital format, and subsequently tested comparatively to establish their relative efficacy. To our knowledge, the contemporary works closest to achieving this goal are guides on how to use the file management features of a specific operating system (e.g., Moran, 2015), but these do not engage with the results of FM or PIM research.

**Mobile FM.** Although some research has examined the problem of the fragmentation of information across devices of varying kinds, and some PIM research has examined tasks like managing contacts on mobile phones (Bergman, Komninos, Liarokapis, & Clarke, 2012), very few works have explicitly examined FM in mobile contexts. When mobile phones (specifically the kind capable of storing files, i.e., smartphones) were first popularised, it was not yet clear if traditional FM would be necessary nor how practical FM would be given the constraints of the device like its screen size, resolution, and input options (i.e., the interactions afforded by mobile devices entail unique challenges; Tungare & Perez-Quinones, 2008). Perhaps as a consequence, end-user file management software was initially scarcely available for mobile phones; early work on mobile FM thus looked to develop prototype mobile FM systems (e.g,. by taking inspiration from records management rather than typical desktop FM; Ballesteros & Moreno, 2007). When mobile FM software was largely unavilable, participants in a study of file synchronisation expressed the desire to have access to the file systems on their mobile devices (Santosa & Wigdor, 2013). Eventually such functionality was provided in default applications on some builds of Android (i.e., depending on the interface overlay) and Apple's iOS (e.g., the Files application on OS 11, iPad build), and is also provided by many third-party applications. To our knowledge only one study has examined mobile file management functionality, finding that file retrieval on phones fails more often and is less efficient than on the computer (Bergman & Yanai, 2017).

It is possible that for many users, cell phones and other mobile devices are used both for casual media consumption and intensive computing; this is suggested, for example, by the plethora of input peripherals for Android devices (e.g., device-specific docking stations with inbuilt keyboards), the availability of mobile versions of office software (e.g., Microsoft Word and Powerpoint), and more generally by device convergence. Consequently, it is possible that on such devices files are being obtained from the Web, email, and cloud storage applications and are stored, edited, backed up, shared, and so on. As the Internet of Things movement continues to grow, mobile FM will likely also expand in scope to include devices beyond mobile phones; for example, if your smart fridge takes a photograph (e.g., of its contents or of your kitchen), is the photo file stored on the fridge? In a folder? Will it later be moved to another device, and then into a folder? Can it be copied or renamed? How will users conceive of and interact with the photo if it is not presented as a file? The less a device resembles a traditional computer, the less straightforward it will be to help users understand and perform FM with that device. Mobile FM is therefore an important direction for future research.

**Improving theory, concepts, methods.**　FM concepts, models, and theories all stand to benefit from refinement in future research. Even the most basic concepts used in FM research can and have reasonably been debated for their precise definitions and general usefulness and vocabulary (Harper et al., 2013): what is a file, what can be done with it, and how should we talk about it? This is no trivial task, as understanding and defining digital objects is very challenging (Hui, 2012), but may be essential if we hope to present clear concepts to FM system users.

The creation of models and theories useful for understanding FM seems a desirable direction for future research. It is yet unclear if existing models or theories could be adapted to FM from broader fields or cognate areas of study; though there is currently no theory available in PIM research, because the basic phenomena of FM may be interesting to many fields (several of which are identified and discussed above) an exploration of the various potential theories seems both possible and valuable. For example, future work may consider adapting information foraging theory (as mentioned above) or and other models from nearby topics (e.g., the records continuum model, as explored by Huvila et al., 2014). Another interesting approach could be to generate theories from the data of FM studies with an inductive methodology (e.g., grounded theory).

We have noted above that most FM studies implicitly employ post-positivist or constructivist epistemologies. Though we do not see obvious problems this might imply for particular studies, differing philosophical assumptions do lend themselves to different interpretations of findings (Hjørland, 2005), and so FM research would likely benefit from careful consideration of this effect. An explication of how different epistemologies would result in different findings in FM research specifically could thus be helpful.

We discussed above the nature and limitations of various approaches to collecting data about users' FM behaviour (i.e., asking, observing, inferring). Dedicated efforts to improve data collection tools may help to overcome such limitations to the benefit of future research. For example, we noted above that the tools currently available for collecting quantitative data (i.e., used to infer user behaviour) do not collect data about many of the available file system properties, and that they are difficult to administer. Should new tools or improvements to existing tools be developed, sharing these for reuse in FM research would benefit the field.

FM research has seen some recent advances in data collection, including the use fMRI technology for observing users during navigation and search tasks (Benn et al., 2015) and dedicated software for collecting file system data (Dinneen, Odoni, Frissen, & Julien, 2016). Future research may further benefit from considering adapting sophisticated methods from HCI and computer science research, like logging and system traces, to record fine-grained data about user behaviour that a file scan does not reveal, like file open times and changes in the size of particular files (M. G. Baker, Hartman, Kupfer, Shirriff, & Ousterhout, 1991; Ousterhout et al., 1985; Roselli, Lorch, & Anderson, 2000). Above we noted that it is desirable to collect longitudinal, or at least time-stamped, FM data (e.g., logging data, file access times, and so on). If this challenge is met, new analytical techniques will be available for analysing and finding insights within such data; for example, methods for tracking information flows across time (e.g., as provided by Luczak-Roesch, Tinati, Van Kleek, & Shadbolt, 2015) may be used to understand the movement of information over time within and across file collections.

**The future of files and FM systems.**   In time, the ideas tested in FM prototypes trickle into both commonly used software and specialised PIM software (Kljun et al., 2015). This fact and the research areas described above might together imply that over the coming years FM software will simply continue to improve incrementally until all FM is performed optimally. But these improvements have come slowly and require a more detailed description of FM behaviour and its component and determining factors than is currently available, and changes in computing initiated by software developers may well modify or replace the FM metaphor before such knowledge is identified. Preliminary conceptions and rumours about such ideas have lead to some common questions (e.g., at scholarly conferences) about the future of file management, including:

- Desktop search is improving and my Mac now comes with support for tagging; won't this solve all of our FM problems and preclude the need for folders?

- I don't organise my music because iTunes does it for me; can't we take the same approach with every file format so that traditional file management becomes unnecessary?

- Organising folders is old fashioned – haven't you heard of *System X*?

We discuss each of these potential future directions in turn.

That search, tagging, or any other feature will replace or preclude the need for folders and organising one's files is an alluring but likely specious hope for the majority of users. Consider the conclusion of the above discussion about search and navigation: though desktop search is undoubtedly useful when navigation fails, folders and navigation aid recognition and reminding more than searching, which lets memories become foggy and thus difficult to recall later. Further established drawbacks to search (e.g., insufficiently distinguishing similar project files) and advantages of folders (e.g., reminding about tasks) are summarised above. Ill-defined information needs, too, are better supported by navigation (or browsing) than by methods requiring the user to explicate that need (Julien et al., 2013) or remember an attribute of an item (e.g., its name or tag), and sense-making is supported by a division of the collection, achieved by folders and reinforced by navigation (W. Jones et al., 2005). Desktop search is powerful, especially when equipped with full-text indexing, but it lacks the dataset that makes Web searching so powerful (e.g., billions of pages and past queries).

Recent work provides further evidence towards this, finding that over a two week period of attempting to perform FM tasks without navigating their folders, some participants were unable to abstain from using folders, later claiming a dependency and implicating folders as essential in PIM task execution and the high-level conceptualisation of their collections (Whitham & Cruickshank, 2017). The previously discussed work by (Benn et al., 2015) provides clues about why folder arrangement may become so ingrained in user behaviour: the human brain has better built-in support for spatial cognition and recognition than for linguistic processing and recall. The likeliness that searching will replace navigating folders is therefore nicely summarised in the paper title *The perfect search is not enough* (Teevan et al., 2004).

In addition to search, other changes in specific computing contexts to how file management is done may lead some to think FM will soon be obsolete, and possibly therefore something that only power users do, like using the command line. This may be motivated by, for example, software that hides file management from the user in favour of managing at the level of a collection or format, like iTunes, as discussed in the introduction. The thought may go as such: why not forego file management entirely, and instead interact with files only when viewed as digital objects of a certain type, in the applications relevant to each type? This is the paradigm, for example, in Apple's mobile operating system, iOS: applications are generally *sandboxed*, or restricted to only seeing files they are responsible for.

It is telling that Apple has not implemented this approach in their desktop OS, and even added a file manager to the iPad in iOS 11. On the desktop, beyond iTunes, the Photos application, and a few other programs files are still interacted with as files, and the Finder file manager application is regularly updated. Arguably, a strict sandbox only moves the general problem of item organisation from the file manager to the format-specific application (e.g., iTunes): items of some format must still be stored, named, organised, assigned metadata, and so on, and once a sufficient number of such items has been stored, it becomes necessary to organise the items with various divisions (i.e., folders or something like them) to facilitate accessing individual items and understanding the whole collection. Interacting with such items without viewing them as files may also be inconvenient; for example, when being sent an email attachment to edit and return, iOS users must download it and hope it appears within the application they intend to use to edit it, and must then push it back to their email from that application. This may be why Google's Chrome OS, despite encouraging the user to do everything in Web-based applications within the Web browser, has a dedicated, if minimal, file manager application.

The guise of avoiding file management is thus lost once interactions beyond basic access to items are desired: when a user wants to send specific songs or photos to another person or device, they may use Dropbox or a USB connection, and will be sending the items as files. But for users of iTunes, this is not a trivial task: the functionality for synchronisation provided by iTunes is to sync an entire collection, songs' file paths are not created by or familiar to the user, and flexible groups of file paths that would have been created by folders (such as an *Artist X* folder) are not readily available. Interestingly, the previously mentioned study by Whitham and Cruickshank (2017), in which participants failed a focused attempt to stop using their folders, took place exclusively in Mac OS. Sandboxing also entails design choices about file type associations, and these are typically motivated by political and commercial desires rather than usability concerns; for example, Apple's music application on iOS does not, without extensive modification, play FLAC format files, and so users must use another application or convert their FLAC files to Apple's proprietary lossless format.

Sandboxing may also make anything beyond lightweight, casual media consumption challenging. For example, in POSIX systems (Mac OS X, GNU/Linux, Unix, BSD; i.e., everything but Windows), everything is regarded as a file – even drives that read removable media. And so, at least for developers, there are too many digital items to not have some abstraction for interacting with, sorting, and accessing them.

Thus, the file and folder paradigm is not easily replaced: there is a need for a common method for interacting with digital items and for organising those items. Sandboxing seems to avoid some of the entailed difficulties of FM, and does so cleverly by drawing on rich file metadata (the sandbox approach works much better for music in standardised formats than for documents), but creates problems of its own for both user interaction (e.g., pushing content) and PIM (e.g., greater fragmentation of a project's files because of their differing formats). The file is a fundamental *cohering* concept between engineers and users that provides a common method for interacting with digital content, and thus "remains central to systems architecture and to the concerns of users" (Harper et al., 2013, p. 1125). Improving upon it therefore likely requires incremental change rather than abandonment: "new abstractions are needed, ones which reflect what users seek to do with their digital data" (Harper et al., 2013, p. 1125).

Finally, in light of the problems identified above in using FMs, it is reasonable to think that a revolutionary idea would be desirable for changing how we interact with digital content, and could come about at any moment. Such revolutions are present in the history of computing interfaces, including for example systems Jef Raskin built on his HCI principles (Raskin, 2000): Canon Cat avoided files by presenting data as a persistent and searchable stream, and Archy added an infinitely zoomable graphical interface. As is common in revolutionary interaction paradigms, though promising, Raskin's projects were
not widely adopted and development eventually ceased. Even as early as the 1960s, the
controversial Project Xanadu (Nelson, 1965) aimed to avoid the paper metaphor by
implementing a hypertext system, which was further revolutionary as it pre-dated the Web
(Nelson, 1965). The original aim of Project Xanadu was to "make a file for writers and
scientists, much like the personal side of Bush's Memex, that would do the things such
people need with the richness they would want... [via] a simple and generalised
building-block structure, user-oriented and wholly general-purpose" (Nelson, 1965, p. 84).
Guided by 17 rules, documents in the Xanadu model contain any kind of digital content
(precluding the need for *files* as such), are linked to other documents based on similar
content, and are intended to be edited while being compared with such items; this is meant
to utilise the digital nature of the documents to support non-sequential authoring and
minimise writing efforts being doubled across documents. Project Xanadu has proven to be
as complex as it is promising, and is still in development; it therefore remains unclear how
the average user, struggling to meet the challenges of classical FM, would feel about using
a *Xanalogical* (Nelson, 1999) interface.

In summary, several incremental and revolutionary prospects show promise for
changing the nature of file management, but the traditional file and folder metaphor also
seems likely to persist. It is tempting to feel that innovative software could at any moment
dramatically change the way we interact with digital objects, and that files will therefore
soon disappear, and yet the existence of equivalent speculation about FM in the 1980s
(e.g., Burton, 1985) suggests this state of uncertainty and promise is not new. Perhaps the
staying power of FM is a consequence of it being a standard, if imperfect, computing
feature, much like the Sholes (i.e., QWERTY) keyboard layout, which has proven difficult
to transition away from. While the above literature shows traditional FM is laden with
challenges, it nonetheless provides an essential computing function by allowing users to
store, organise, retrieve, share, and interact with many digital items, and FM and its
challenges both seem to make for an interesting and productive object of study. Finally, as
noted above, many novel and potentially beneficial FM augmentations and interfaces are
never adopted by users or integrated into default FM applications. Increasing the use of
such software, perhaps through more direct engagement with OS vendors, is therefore an
outstanding long-term goal for the PIM and HCI research communities.

## Conclusion

File management is a ubiquitous and challenging activity. In this manuscript we have
synthesised works from various disciplines examining this activity, and have identified that such work typically aims to understand users' FM behaviour, the factors determining it, and how these results can be used to improve the relevant systems and services. Such works have been undertaken by researchers active in information science, personal information management, human-computer interaction, computer science, and more, and have drawn upon various methods from these fields; the study of FM is thus interdisciplinary and potentially highly impactful for these fields and those with overlapping interests, such as psychology and information visualisation, retrieval, and organisation. This is perhaps unsurprising, given the apparent fundamental nature of the file and folder context, where users manage items in bespoke information structures.

What the study of FM faces in the future is a shifting landscape where user behaviour is difficult to study, analyse, and support, because it is nuanced, private, personal, and changing along with its technological context. The implications of increases in use of the cloud, available storage space, fragmentation of information across devices, and complex social information management on FM are unclear. Several advances are needed to better understand and support users' FM activities: new data collection tools, further studies, models of user behaviour, and theories. Through such advances and the findings of FM studies (like those presented above), FM research holds connections to and implications for other areas of study surrounding information behaviour and structures, and shows promise for improving the ubiquitous task of file management through the informed designed of new software and services.

## Acknowledgements

The authors thank Jamshid Beheshti, Rob Capra, Ilja Frissen, Ben Hanrahan, Maja Krtalić, and Shane McIntyre for their comments on drafts of this manuscript. The authors also thank Lauren Bennett and Rodreck David for their help in preparing the manuscript for submission.